\documentclass[preprint,showpacs,showkeys,preprintnumbers, amsmath,amssymb, aps, prfluids,floatfix,superscriptaddress]{revtex4-2}

\usepackage{color}
\usepackage{amsmath}
\usepackage{graphicx}
\usepackage{dcolumn}
\usepackage{bm}
\usepackage{hyperref}
\usepackage{etoolbox}

\usepackage{soul}
\usepackage{lmodern}

\newcommand{\bq}{\boldsymbol{q}}

\newcommand{\bu}{\boldsymbol{u}}

\newcommand{\bI}{\boldsymbol{I}}

\begin{document}

\title{
Thermal Enskog-Vlasov Lattice Boltzmann model with phase separation
}

\author{Sergiu Busuioc}
\email[]{sergiu.busuioc@e-uvt.ro}
\affiliation{Department of Physics, West University of Timişoara\\
Bd. Vasile Pârvan 4, 300223 Timişoara, Romania}
\affiliation{Institute for Advanced Environmental Research, West University of Timişoara\\
Bd. Vasile Pârvan 4, 300223 Timişoara, Romania}
\author{Victor Sofonea}
\email[]{sofonea@gmail.com}
\affiliation{Department of Physics, West University of Timişoara\\
Bd. Vasile Pârvan 4, 300223 Timişoara, Romania}
\affiliation{Center for Fundamental and Advanced Technical Research, Romanian Academy\\
Bd. Mihai Viteazul 24, 300223, Timişoara, Romania}
{\large }
\date{\today}

\begin{abstract}
An Enskog-Vlasov finite-difference Lattice Boltzmann (EV-FDLB) for liquid-vapor systems with variable temperature is introduced.  The model involves both the simplified Enskog collision operator and the self-consistent force field which accounts for the long-range interaction between the fluid particles.  Full-range Gauss-Hermite quadratures were used for the discretization of the momentum space.
The numerical solutions of the Enskog-Vlasov equation obtained employing the EV-FDLB model and the Direct Simulation Monte Carlo (DSMC)-like particle method (PM) are compared. Reasonable agreement is found between the two approaches when simulating the liquid-vapor phase separation and the liquid slab evaporation.
\end{abstract}

\keywords{ Enskog-Vlasov equation; liquid-vapor systems, lattice Boltzmann; Discrete Simulation Monte Carlo.  }

\maketitle


\section{\label{sec:intro}Introduction}

In past decades, considerable attention was paid to the use of the Boltzmann equation for the investigation of rarefied gases and micro-scale flow phenomena, where the value of the Knudsen number (the ratio between the mean free path of fluid particles and the characteristic size of the flow domain) is no longer negligible.  In Boltzmann's theory, the fluid constituents are considered to be point particles, i.e. their size is negligible when compared to their mean free path. This conjecture is declined in the case of dense fluids, where the Enskog equation is used instead. Unlike the Boltzmann equation, the Enskog equation takes into account the space correlations between colliding particles, the mutual shielding, as well as the reduction of the available volume \cite{cowling70,FK72,DBK2021}.

Similarly to the Boltzmann equation, the  Enskog equation may be solved numerically using the deterministic or the probabilistic methods existing in the current literature. When investigating fluid flow problems using these kinetic equations, particular interest was given to the widely used direct simulation Monte Carlo (DSMC) method \cite{B76, B13, BDSMC,  ARS23}. A particle method (PM) that extends the DSMC method to handle the Enskog collision term in computer simulations was devised by Frezzotti \cite{F97b}. Later, a self-consistent force was added to the Enskog equation to account for the attractive part of the long-range interparticle interactions, giving the commonly referred Enskog-Vlasov (EV) equation \cite{S67, G71, KS81, HD02, FGL05, BB19}. The EV equation has been widely applied to study two-phase flows and liquid-vapor phase transitions \cite{DBK2021, FGL05, TMHH18, BB18, takata2018waals, BB19, FBG19, BGLS20, BG20, takata2021kinetic, BFG23}.  In such investigations, computer simulations with particle models (PMs) incorporating long-range interparticle forces play an important role. The considerable computational resources demanded by PMs pose significant limitations in the case of large-scale problems, as well as in engineering design activities. Consequently,  kinetic model equations that reduce the evaluation costs of both the Enskog collision operator and the self-consistent interparticle force become highly desirable. To address this issue, a common procedure is to simplify the Enskog collision integral by expanding it into a Taylor series, resulting in the so-called simplified Enskog collision operator. Recently, the simplified Enskog collision operator has been successfully integrated into various solvers, including the discrete unified gas kinetic scheme (DUGKS)~\cite{CWWC23}, the discrete velocity method (DVM)~\cite{WWHLLZ20,WGLBZ23,LSSLGZ24}, the discrete Boltzmann method (DBM)~\cite{ZXQWW20, GXLLSS22}, the double-distribution Lattice Boltzmann model (DDLB)~\cite{HWA21}, and the finite-difference Lattice Boltzmann (FDLB) models~\cite{B23,BS24,B24}. These solvers provide computationally efficient alternatives for studying micro-scale flow phenomena while maintaining reasonable accuracy.

In this paper, we further extend the Enskog FDLB models proposed in Refs. \cite{B23,BS24,B24}, by incorporating the long-range molecular attraction using the mean-field approximation. This adds a weak attractive tail to the force term in the Enskog equation, which eventually generates the formation of liquid-vapor interfaces in the flow domain. The resulting Enskog-Vlasov FDLB (EV-FDLB) model agrees with the kinetic model of He and Doolen \cite{HD02}.
The computer simulations reported in this paper were conducted with our EV-FDLB model using a full-range Gauss-Hermite quadrature involving high-order velocity sets \cite{SYC06, AS16a, AS16b, AS2019, BA19, B23, BS24, B24}. These velocity sets are off-lattice, hence, finite difference schemes were used in this paper to solve the model evolution equations numerically.

Due to the complex structure of the mean-force field integral, its direct numerical evaluation is expensive.  An alternative procedure involves the approximation of this integral by assuming that the density is a smooth and slowly varying function everywhere in the flow domain, including the phase interfaces. Throughout this paper, these two procedures are denoted by EV1 and EV2, respectively. One of the objectives of this study is to test the behavior of the EV2 approximation by comparison to the EV1 version. Furthermore, the simulation results obtained with the two distinct force field evaluations are benchmarked against the corresponding results supplied by the particle method (PM) described in \cite{FGL05,FBG19,BGLS20} in the case of two thermal flow problems.

The paper is organized as follows. In Section \ref{sec2} we introduce the EV-FDLB model and describe the simplified Enskog collision operator, as well as the two procedures (EV1 and EV2) used to numerically evaluate the self-consistent force field. In Section \ref{sec3} we compare {\color{blue}the results for the liquid-vapor interface profile and the phase diagram of an isothermal EV fluid modelled using the PM and the two EV-FDLB procedures}. Furthermore, the EV-FDLB results are benchmarked with respect to the PM results by considering two test problems with variable temperature: a - the phase separation dynamics in an EV fluid with variable temperature; b -  the evaporation of a liquid slab surrounded by metastable vapor.
We conclude the paper in Section \ref{sec4}. The Appendix \ref{app:PM} provides a brief outline of the PM used to validate the EV-FDLB simulations, while the Appendix \ref{app:runtime} lists the computational time required by each method.

\section{Enskog-Vlasov thermal model }\label{sec2}

\subsection{Enskog-Vlasov equation}

Consider a fluid comprised of spherical particles, all identical with mass $m$ and diameter $\sigma$, interacting through the Sutherland potential which combines a hard sphere potential with an attractive soft potential tail\cite{cowling70}:
\begin{equation}\label{eq:sutherland}
 \phi(\varrho)=
 \begin{cases}
   +\infty \,,\quad   &\varrho < \sigma, \\
   \displaystyle{-\phi_\sigma\left( \frac{\varrho}{\sigma}\right)^{-\gamma}}\, ,\quad &\varrho\geq \sigma,
  \end{cases}
\end{equation}
Here $\varrho=||\bm{r}_1-\bm{r}||$ represents the distance between two interacting particles located at positions $\bm{r}_1$ and $\bm{r}$, while the positive constants $\phi_\sigma$, $\gamma$ define the depth of the potential well and the range of the soft interaction, respectively.

The exact evolution equation for the one-particle distribution function $f(\bm{r},\bm{p},t)$  describing this fluid was derived by Karkheck and Stell in 1981 \cite{KS81}. However, this equation involves the two-particle distribution function and is of limited utility. Two simplifying assumptions may be considered together in order to obtain a closed equation: the neglection of the long-range particle correlations and the approximation of the short-range particle correlations using the Enskog theory initially developed for dense gases. With these assumptions, one gets the following closed equation governing the evolution of the one-particle distribution function $f\equiv f(\bm{r},\bm{p},t)$\cite{G71,FGL05}:
 \begin{equation} \label{eq:ev_general}
  \frac{\partial f}{\partial t}
  +\frac{\bm{p}}{m}\cdot\nabla_{\bm{r}}f + \frac{{\bm{\mathcal{F}}}[n]}{m} \cdot \nabla_{\bm{p}}f=J_E[f],
 \end{equation}
where $\bm{p}$ is the particle momentum, $n \equiv n(\bm{r},t)$ is the local number density and the square brackets  denote functional dependence.
In Eq.~\eqref{eq:ev_general} above,
\begin{equation}\label{eq:force}
 \bm{\mathcal{F}}[n] =  \int_{\varrho>\sigma} \frac{d\phi(\varrho
 )}{d\varrho}\frac{\bm{r}_1-\bm{r}}{||\bm{r}_1-\bm{r}||}n(\bm{r_1})d\bm{r_1}
 \end{equation}
is  the self-consistent force field generated by the soft attractive tail and the hard-sphere collision integral $J_E\equiv J_E[f]$ is defined by:
\begin{multline}
 J_E = \sigma^2\int(\bm{p}_r\cdot\bm{\hat{k}})^+d\bm{p}_1d^2\bm{\hat{k}}
\left\{\chi\left[n\left(\bm{r}+\frac{\sigma}{2}\bm{\hat{k}},t\right)\right] f(\bm{r}+\sigma\bm{\hat{k}},\bm{p}_1^*,t) f(\bm{r},\bm{p}^*,t)-\right.\\
  \left.\chi\left[n\left(\bm{r}-\frac{\sigma}{2}\bm{\hat{k}},t\right)\right] f(\bm{r}-\sigma\bm{\hat{k}},\bm{p}_1,t) f(\bm{r},\bm{p},t)  \right\},\label{eq:enskog}
\end{multline}
where $\bm{p_r}=\bm{p_1}-\bm{p}$ is the relative momentum,  $\bm{\hat{k}}$ is the unit vector giving the relative positions of two colliding particles,
$(\cdot)^{+}$ indicates that the surface integral is restricted to the half-sphere satisfying $\bm{p}_r\cdot\bm{\hat{k}}>0$ and $\chi[n]$ is the contact value of the pair correlation function in a hard-sphere fluid at equilibrium.
In the Standard Enskog Theory (SET), {\color{blue} {$\chi[n]$ is replaced by the pair correlation function $\chi_{\mbox{\tiny SET}} [n]$}} evaluated in a fluid at uniform equilibrium, where the number density $n$ is  calculated at the contact point of the two colliding molecules. An approximate but accurate expression for $\chi_{\mbox{\tiny SET}} [n]$, which was derived from the Carnahan-Starling equation of state for hard-sphere fluid~\cite{CS69} is given by:
\begin{equation}
\chi_{\mbox{\tiny SET}}[n] =\frac{1}{nb}\left(\frac{p^{hs}}{n k_B T}-1\right)=\frac{1}{2}\frac{2-\eta}{(1-\eta)^3},
\qquad b=\frac{2\pi \sigma^3}{3},\qquad \eta=\frac{\pi \sigma^3 n}{6}
\label{eq:carnstar}.
\end{equation}
where $p^{hs}$ denotes the pressure of a system of hard spheres and $k_B$ represents the Boltzmann constant.
In this work, {\color{blue} {we follow the Fischer-Methfessel approach~\cite{FM80} and replace  the density $n$  at the contact point }} with the density field $\overline{n}(\bm{r},t)$ averaged over a spherical volume of radius $\sigma$\,:
\begin{subequations}
\begin{equation}
\chi\left[n\left({\bm{r}}\pm\frac{\sigma}{2}\bm{k},t\right)\right]=\chi_{\mbox{\tiny SET}}\left[\overline{n}\left({\bm{r}}\pm\frac{\sigma}{2}\bm{k},t\right)\right],
\end{equation}
where
\begin{equation}
\overline{n}(\bm{r},t)=\frac{3}{4\pi \sigma^3}\int_{\mathbb{R}^3} n(\bm{r}_*,t)w(\bm{r},\bm{r}_*)\,d\bm{r}_*, \hspace{1cm}
w(\bm{r},\bm{r}_*)=\left\{
\begin{array}{cc}
1, &\qquad \|\bm{r}_*-\bm{r}\|<\sigma, \\
0, & \qquad \|\bm{r}_*-\bm{r}\|>\sigma.
\end{array}
\right.
\end{equation}
\end{subequations}
{\color{blue} {For convenience, in the rest of this paper we write $\chi$ instead of $\chi_{\mbox{\tiny SET}}\left[\overline{n}\right]$.}}

The kinetic equation~\eqref{eq:ev_general} is commonly known as the Enskog-Vlasov (EV) equation~\cite{S67,G71,KS81,FGL05,BB19}. The EV equation has been utilized to investigate a variety of two-phase flows, including monoatomic fluids \cite{FGL05, KKW14, FBG19, BGLS20}, polyatomic fluids \cite{Bruno2019, BG20}, mixtures \cite{KSKFW17}, the formation and rupture of liquid menisci in nanochannels \cite{BFG15}, as well as the growth and collapse dynamics of spherical nano-droplets and bubbles \cite{BFG23}.

\subsection{Simplified Enskog collision operator}

Assuming smoothness of both the pair correlation function $\chi$ and the distribution functions $\{f^*\equiv f({\bm{x}},\bm{p^*},t),f_1^*\equiv f({\bm{x}} + \sigma {\bm{k}},\bm{p_1^*},t),f\equiv f({\bm{x}} ,\bm{p},t) ,f_1\equiv f({\bm{x}} - \sigma {\bm{k}},\bm{p_1},t)\}$, we can approximate these functions in the Enskog collision integral $J_E$ through a Taylor series around the point $\bm{x}$. The resulting terms up to first-order gradients are \cite{cowling70,K10,WWHLLZ20,WGLBZ23}:
\begin{eqnarray}
J_E&\simeq& J_0+J_1\label{eq:joplusj1}\\
 J_0 \equiv\,   J_0[f] &=& \chi\int (f^*f_1^*-ff_1)\sigma^2({\bm{p_r}}\cdot{\bm{k}}) d{\bm{k}}d{\bm{p_1}}\\
 J_1 \equiv\,   J_1[f] &=& \chi\sigma\int\bm{k}(f^*\bm{\nabla}f_1^*-f\bm{\nabla} f_1)\sigma^2({\bm{p_r}}\cdot{\bm{k}}) d{\bm{k}}d{\bm{p_1}}\nonumber\\
 &+&\frac{\sigma}{2}\int\bm{k}\bm{\nabla}\chi(f^*f_1^*-ff_1)\sigma^2({\bm{p_r}}\cdot{\bm{k}}) d{\bm{k}}d{\bm{p_1}}
\end{eqnarray}
where all functions $f^*,f_1^*,f,f_1$, and $\chi$ are evaluated at the point $\bm{x}$.

The collision term $J_0$ resembles the conventional collision term of the Boltzmann equation multiplied by $\chi$ and is treated accordingly by employing the relaxation time approximation. In this context, we utilize the Shakhov collision term \cite{shakhov68a,shakhov68b}:
\begin{equation}
 J_0=-\frac{\chi}{\tau}(f-f^S),
\end{equation}
where $\tau$ {\color{blue}{is the relaxation time given in Eq. \eqref{eq:tau} below}} and $f^S$ denotes the equilibrium Maxwell-Boltzmann distribution:
\begin{equation}
 f_{\text{\tiny MB}}=\frac{n}{(2 m \pi k_B T)^{3/2}}\exp{\left(-\frac{\bm{\xi}^2}{2 m k_B T}\right)}
\end{equation}
multiplied by a correction factor \cite{shakhov68a,shakhov68b,GP09,ASS20}:
\begin{equation}
 f^S=f_{\text{\tiny MB}}\left[1 + \frac{1-\text{Pr}}{P_i k_B T}\left( \frac{\bm{\xi}^2}{5 m k_B T}-1 \right)\bm{\xi}\cdot \bm{q} \right].
\end{equation}
Here $\bm{q}$ represents the heat flux calculated as:
\begin{equation}
 \bm{q}=\int d^3p f \frac{\bm{\xi}^2}{2m}\frac{\bm{\xi}}{m},
\end{equation}
$\bm{\xi}=\bm{p}-m\bm{u}$ denotes the peculiar momentum, $\text{Pr}=c_P\mu/\lambda$ stands for the Prandtl number, $c_P=5k_B/2m$ represents the specific heat at constant pressure, and $P_i=\rho R T=n k_B T$ signifies the ideal gas equation of state, with $R\,=k_{B}/m$ being the specific gas constant.
It's worth noting that although the Shakhov model doesn't guarantee the non-negativity of the correction factor and the proof of the H-theorem is still pending, the model has been validated through experimental \cite{S02,S03,GP09} or DSMC \cite{AS18,ZXZCW19,ASS20,TWS20} results.

The second term of $J_E$, denoted as $J_1[f]$, can be approximated by replacing the distribution functions ($f^*,f_1^*,f,f_1$) with the corresponding equilibrium distribution functions. By employing $f_{\text{\tiny MB}}^*f_{\text{\tiny MB},1}^*=f_{\text{\tiny MB}}f_{\text{\tiny MB},1}$ and integrating over $\bm{k}$ and $\bm{p_1}$, one gets \cite{cowling70,K10,WWHLLZ20,WGLBZ23,SF22}:
\begin{multline}
 J_1[f]\simeq J_1(f_{\text{\tiny MB}},f_{\text{\tiny MB}})=\\-b \rho \chi f_{\text{\tiny MB}} \left\{\bm{\xi}\left[\bm{\nabla}\ln(\rho^2 \chi T)+\frac{3}{5}\left(\zeta^2-\frac{5}{2}\right)
 \bm{\nabla}\ln T\right]\right.\\
 \left. + \frac{2}{5}\left[ 2\bm{\zeta}\bm{\zeta}\bm{:\nabla u} + \left(\zeta^2-\frac{5}{2} \right)\bm{\nabla\cdot u} \right]
 \right\}
\end{multline}\label{eq:J1}
where $\bm{\zeta}=\bm{\xi}/\sqrt{2 m k_B T}$. With the aforementioned approximations and assuming no external force, the Enskog equation Eq.~\eqref{eq:ev_general} can be simplified to:
\begin{equation}\label{eq:enskog_approx}
 \frac{\partial f}{\partial t}+\frac{\bm{p}}{m}\nabla_{\bm{x}}f+\frac{ \bm{\mathcal{F}}[n]}{m} \cdot \nabla_{\bm{p}}f=-\frac{\chi}{\tau}(f-f^S)+J_1(f_{\text{\tiny MB}},f_{\text{\tiny MB}})
\end{equation}

The macroscopic quantities are evaluated as moments of the distribution function:
\begingroup
\renewcommand*{\arraystretch}{1.25}

\begin{equation}
 \begin{pmatrix}
 n \\ \rho\bm{u} \\ \frac{3}{2} n k_B T
 \end{pmatrix} =\int d^3 p
 \begin{pmatrix}
 1 \\ \bm{p} \\ \frac{\bm{\xi}^2}{2m}
 \end{pmatrix} f
\end{equation}
\endgroup
where $\rho=mn$.

\subsection{Self-consistent force field evaluation }

In the EV-FDLB simulations presented in this paper, the numerical evaluation of the self-consistent force field in the EV equation \eqref{eq:ev_general} is performed in two ways.

The first one involves the direct numerical integration in \eqref{eq:force} using the composite Simpson's 1/3 rule. As shown in Ref.\cite{FGL05}, in the one-dimensional case the following simplified expression for the $x$ component of the self-consistent force field can be derived:
\begin{equation}\label{eq:F1}
 \bm{\mathcal{F}}_1[n]=2\pi\phi_\sigma\left[ \sigma^\gamma\int_{|x-x'|>\sigma} \frac{(x'-x)n(x',t)}{|x-x'|^\gamma}dx' + \int_{|x-x'|<\sigma} (x'-x)n(x',t)dx'\right]
\end{equation}

The second one assumes that the number density $n$ is sufficiently smooth everywhere, including the phase interface. This enables the expansion of the number density in Eq.~\eqref{eq:force} to get the following approximation of the Vlasov force \cite{SF22}:
\begin{equation}\label{eq:F2}
 \bm{\mathcal{F}}_2[n]=2a\nabla n + \kappa \nabla\Delta n=
 \frac{4\pi\sigma^3}{3}\phi_\sigma\frac{\gamma}{\gamma-3}\nabla n + \frac{2\pi \sigma^5}{15}\phi_\sigma\frac{\gamma}{\gamma-5}\nabla\Delta n,
\end{equation}
where $a=\frac{2\pi\sigma^3}{3}\phi_\sigma\frac{\gamma}{\gamma-3}$ and $\kappa=\frac{2\pi \sigma^5}{15}\phi_\sigma\frac{\gamma}{\gamma-5}$.

 Thus, in the FDLB framework we have two versions of the evolution equations, denoted EV1 and EV2:
\begin{subequations}\label{eq:ev_12}
 \begin{align}
\text{EV1 :}\quad\frac{\partial f}{\partial t} + \frac{\bm{p}}{m}\cdot\nabla_{\bm{r}}f+\frac{\bm{\mathcal{F}}_1[n]}{m} \cdot \nabla_{\bm{p}}f&=J_0[f]+J_1[f],\label{eq:ev1}\\
\text{EV2 :}\quad\frac{\partial f}{\partial t} + \frac{\bm{p}}{m}\cdot\nabla_{\bm{r}}f+\frac{\bm{\mathcal{F}}_2[n]}{m} \cdot \nabla_{\bm{p}}f&=J_0[f]+J_1[f]\label{eq:ev2}
 \end{align}
 \end{subequations}

The equation of state of the fluid described by the Enskog-Vlasov equation has a generalized van der Waals form \cite{B75}:
\begin{equation}\label{eq:EOSS}
  P(n,T)=nkT \frac{1+\eta+\eta^2-\eta^3}{\left(1-\eta\right)^3} -a n^2
\end{equation}
where the repulsive part is given by the Carnahan Starling equation of state.

The Enskog-Vlasov equation leads to the following fluid dynamics equations\cite{HD02}:
\begin{align}
     & \partial_t \left( \rho \right) + \nabla \cdot \left( \rho \bu \right) = 0 , \\
     & \partial_t \left( \rho \bu \right) + \nabla \cdot \left( \rho \bu \bu +\mathbf{\Pi} \right) = 0 , \\
     & \partial_t  E +\nabla \cdot(\bu E) + \mathbf{\Pi}:\nabla \bu +
       \nabla \bq -\kappa\nabla\bu:[\nabla(\rho\nabla\rho)-\frac{1}{2}\nabla\cdot(\rho\nabla\rho)\bI]= 0,
\end{align}
where $E$ is the total internal energy (kinetic and potential). The expressions of the stress tensor and heat flux are \cite{HD02}:
\begin{align}\label{eq:heatstress}
    \mathbf{\Pi} & = \left(P  -\kappa \rho \nabla^2 \rho - \frac{\kappa}{2} |\nabla \rho|^2\right) \bI - \mu \left( \nabla \bu + \nabla \bu^{\mbox{\tiny T}} \right) + \kappa \nabla \rho \nabla \rho, \\
    \mathbf{q}  & = - \lambda \nabla T.
\end{align}
where $\mathcal{I}$ represents the identity matrix.The shear viscosity $\mu$ and the thermal conductivity $\lambda$, which appear in Eqs.~{\eqref{eq:heatstress}, are given by~\cite{K10}:
\begin{eqnarray}
 \mu & = & \mu_0 \left[\frac{1}{\chi}+\frac{4}{5}(b\rho)+\frac{4}{25}\left(1+\frac{12}{\pi}\right)(b\rho)^2\chi\right]\, , \label{eq:viscosity} \\
 \lambda & = & \lambda_0 \left[\frac{1}{\chi}+\frac{6}{5}(b\rho)+\frac{9}{25}\left(1+\frac{32}{9\pi}\right)(b\rho)^2\chi\right]\, .
\end{eqnarray}
In these equations, $\mu_0$ and $\lambda_0$ represent the viscosity and the thermal conductivity for hard-sphere molecules at temperature $T$, namely~\cite{K10}:
\begin{equation}
\mu_0=\frac{5}{16\sigma^2}\sqrt{\frac{m k_B T}{\pi}},\quad \lambda_0=\frac{75 k_B}{64 m\sigma^2}\sqrt{\frac{m k_B T}{\pi}}.
\end{equation}
The Chapman-Enskog expansion of Eq.~\eqref{eq:enskog_approx} provides the relationships between the relaxation time $\tau$ and the transport coefficients. In this context, the relaxation time $\tau$ is expressed as\cite{HD02}:

\begin{equation}\label{eq:tau}
\tau=\frac{\mu}{P_i(1+\frac{2}{5}b\rho\chi)}=\frac{\lambda}{\frac{5k_B}{2m}\frac{\tau P_i}{\text{Pr}}(1+\frac{3}{5}b\rho\chi)}
\end{equation}
resulting in the following expression for the Prandtl number:
\begin{equation}\label{eq:prandtl}
\text{Pr}=\frac{2}{3} \, \frac{(1+\frac{3}{5}b\rho\chi)}{(1+\frac{2}{5}b\rho\chi)} \,\frac{1+\frac{4}{5}b\rho\chi+\frac{4}{25}\left(1+\frac{12}{\pi}\right)(b\rho\chi)^2}{1+\frac{6}{5}b\rho\chi+\frac{9}{25}\left(1+\frac{32}{9\pi}\right)(b\rho\chi)^2}.
\end{equation}

The model equations involving the simplified Enskog collision operator $J_1$ use only a limited number of low-order derivatives, omitting information contained in higher-order terms. Consequently, the high-order information excluded in $J_1$, which does not appear in the collisional momentum and energy transfer, is reintroduced in the kinetic transfer of momentum and energy through the relaxation time \eqref{eq:tau} and the Prandtl number \eqref{eq:prandtl} in the collision term $J_0[f]$. This reintroduction ensures that the total stress tensor and heat flux derived from the current kinetic model align with those obtained from the Enskog equation, at least up to the first-order approximation\cite{K10,WGLBZ23}.

\section{Comparison between EV-FDLB and PM results}\label{sec3}

Throughout the paper, we follow the non-dimensionalization procedure outlined in Ref.~\cite{AS18,NBAGLS19}, which involves three reference quantities: $L_{\text{ref}}$ (length), $n_{\text{ref}}$ (particle number density), and $T_{\text{ref}}$ (temperature). Consequently, the reference momentum is defined as $p_{\text{ref}} = \sqrt{m_{\text{ref}} k_B T_{\text{ref}}}$, and the reference time is given by $t_{\text{ref}} = \frac{m_{\text{ref}} L_{\text{ref}}}{p_{\text{ref}}}$, where $m_{\text{ref}}$ denotes the mass of a fluid particle.

For convenience, in our simulations, the non-dimensionalized value $\sigma$ of the particle diameter is set to 1. Moreover, we select the values $\phi_\sigma = 1$ and $\gamma = 6$ to ensure the same far-field behavior as the 12-6 Lennard-Jones potential~\cite{HBC64}. The constants $\sigma$, $\phi_\sigma$ and $\gamma$ appear in the expression \eqref{eq:sutherland} of Sutherland's potential, as well as in the approximation  \eqref{eq:F2} of the Vlasov force.

In this paper, we assume that the EV fluid is homogeneous along both the $y$ and the $z$ directions of the Cartesian coordinate system. This allows us to reduce the three-dimensional molecular velocity space to a one-dimensional one by employing reduced distribution functions, as discussed in Refs. \cite{WWHLLZ20,WGLBZ23,B23,BS24}. The one-dimensional versions of the kinetic evolution equations \eqref{eq:ev_12} are solved using the finite-difference Lattice Boltzmann method with full-range Gauss-Hermite quadratures \cite{ABWPK19,SBBAGL18,BABS20,ASS20,B23,BS24,B24}. The numerical schemes used were: the third-order total variation diminishing (TVD) Runge-Kutta method for time-stepping\cite{SO88}, the fifth-order WENO-5 advection scheme \cite{GXZL11,JS96}, the $4$th order central difference scheme used for gradient evaluation\cite{F88}, and the 5 point stencil in Ref.~\cite{PK06} for the gradient of the Laplacian appearing in Eq.\eqref{eq:F2}. The EV-FDLB simulations were conducted with the time step $\delta t=5\times10^{-3}$,  the lattice spacing $\delta x=\sigma/10$ and the quadrature order $Q_x=16$.

The numerical results obtained with the  EV1 and the EV2 versions of our EV-FDLB code are compared systematically with the results obtained using a DSMC-like particle method (PM) that extends the original Direct Simulation Monte-Carlo (DSMC) method to handle the nonlocal nature of the Enskog collision integral \eqref{eq:enskog}.  Details regarding the PM can be found in Appendix \ref{app:PM}.

The main goal of this study is to establish the degree to which the EV2 version of the code, which involves the approximation \eqref{eq:F2} of the self-consistent force \eqref{eq:force}, is impacting the simulation results when investigating flow problems with phase change. In this regard, we first discuss the stationary liquid-vapor interface profile, as traced in isothermal conditions with the {\color{blue} {PM, as well as the EV1 and EV2}} versions, for various values of the fluid temperature. Then, we consider two numerical experiments where the temperature changes dynamically.

\subsection{Interface profile and phase diagram}\label{ssec:phasediag}

\begin{figure}
\includegraphics[width=0.85\linewidth]{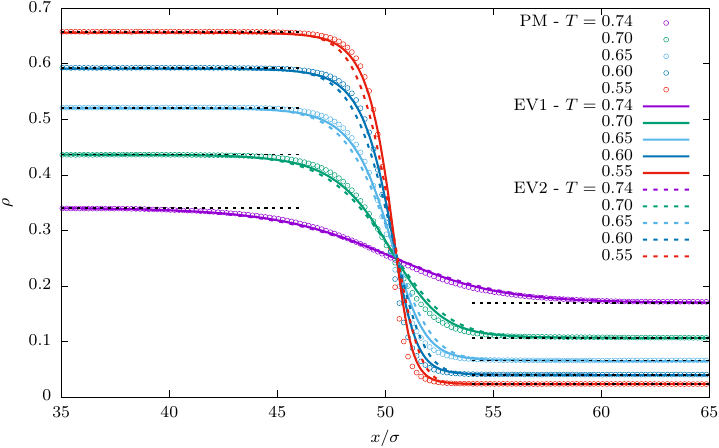}
\caption{Density profiles at the liquid-vapor interface, obtained using both the PM and the FDLB methods (EV1 and EV2) at 5 {\color{blue} {constant}} values of the temperature $T$. For each $T$, the {\color{blue} {horizontal}}  black-dotted lines indicate the values of both the liquid and the vapor density, as calculated according to the Maxwell construction. \label{fig:interface}}\end{figure}

In the first test, the PM and the two FDLB versions (EV1 and EV2) of the code were used to investigate the stationary density profiles across planar liquid-vapor interfaces established in EV fluids at constant temperatures. The simulations were conducted with five values of the temperature, namely $T\in\{0.55,0.6,0.65,0.7,0.74\}$, which are below the critical temperature value $T_c \simeq 0.754632$ of the EV fluid \cite{FGL05}.  At the beginning of each  simulation, the initial  profile $\rho_{i}(x)$ was set according to:
\begin{equation}\label{eq:tanh}
 \rho_{i}(x)=\rho_\ell + \frac{\rho_v - \rho_\ell }{ 2} \left[1 + \tanh\left(\frac{|x| - x_0 }{ \xi}\right)\right].
\end{equation}
where  $\rho_\ell$ and $\rho_v$ are the liquid and vapour density calculated at temperature $T$ using the Maxwell construction, $\xi=\sqrt{ \frac{3. }{ 4 \pi \sqrt{3}  (T_c - T)}}$ and $x_0=50\sigma$. In all simulations reported in this subsection, the {\color{blue} {fluid}} temperature $T$ was kept constant. The computational domain size was $[-100\sigma:100\sigma]$, with periodic boundary conditions along the $x$ axis. A number of $10^6$ iterations were always performed to ensure the quasi-stationary profile for all temperatures $T$.

\begin{figure}
\includegraphics{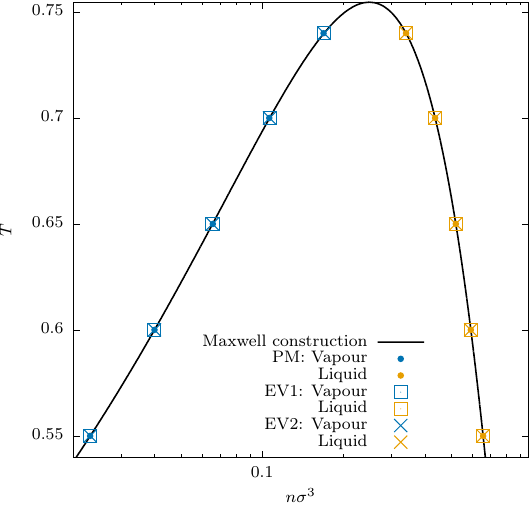}
\caption{ Phase diagram (fluid density $\rho = n\sigma^3$ vs temperature  $T$) : the continuous line was traced according to the Maxwell construction for the equation of state \eqref{eq:EOSS}, while the symbols denote the density values calculated numerically using the PM, as well as the EV1 and the EV2 procedures.}
\label{fig:phase_diagram}
\end{figure}

\begin{figure}
\begin{tabular}{cc}
\includegraphics[width=0.43\linewidth]{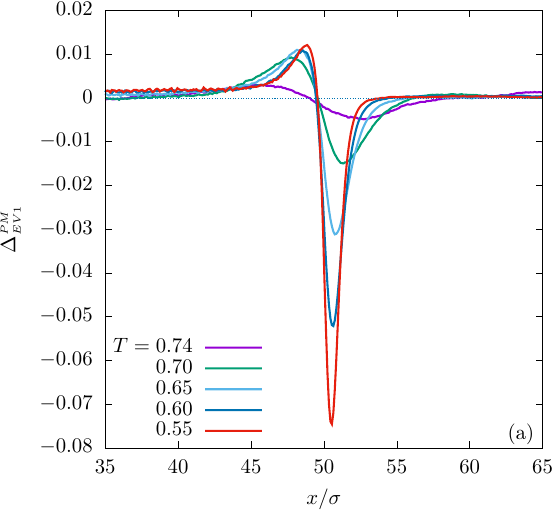}&
\includegraphics[width=0.42\linewidth]{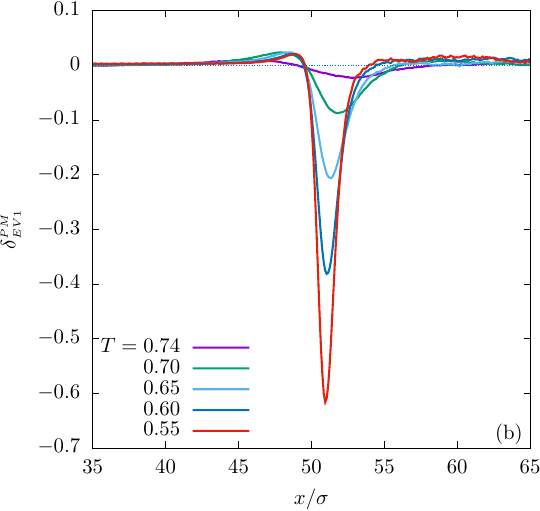}\\
\includegraphics[width=0.43\linewidth]{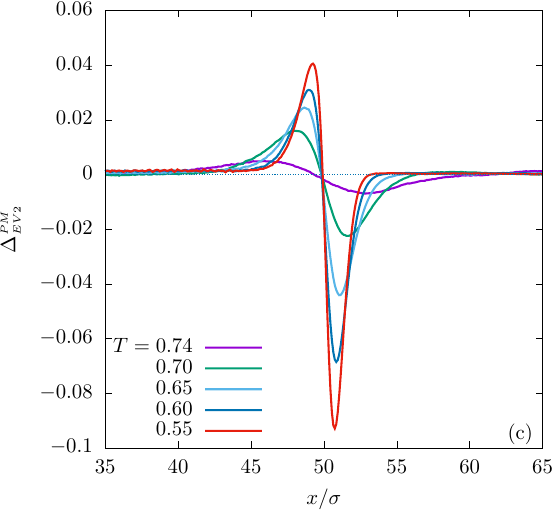}&
\includegraphics[width=0.42\linewidth]{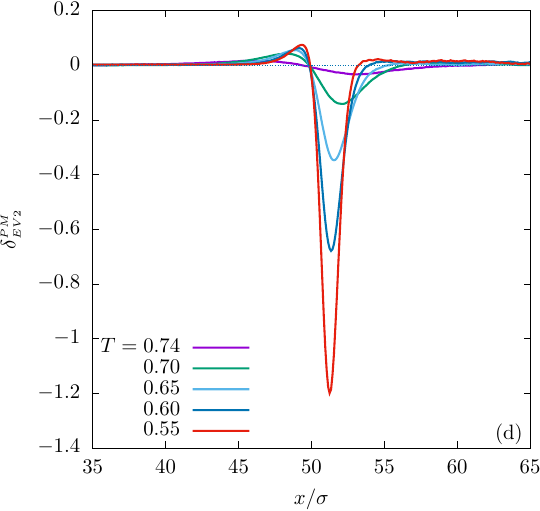}\\
\includegraphics[width=0.43\linewidth]{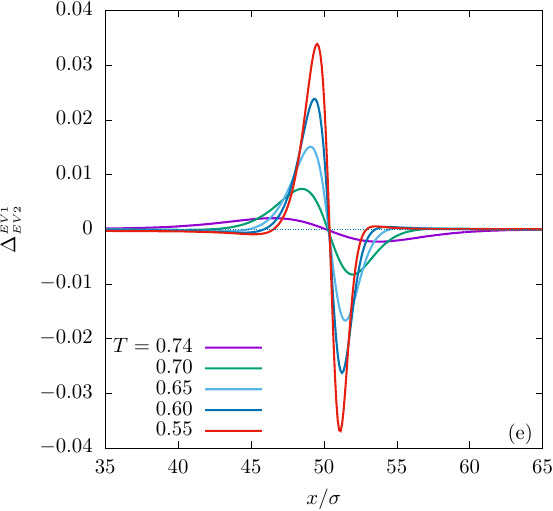}&
\includegraphics[width=0.42\linewidth]{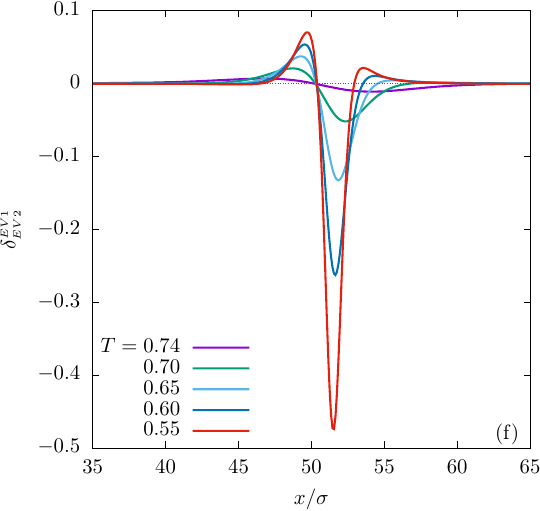}
\end{tabular}
\caption{\color{blue} The local errors $\Delta^{\text{\tiny PM}}_{\text{\tiny EV1}}$  (a), $\Delta^{\text{\tiny PM}}_{\text{\tiny EV1}}$  (c)   $\Delta^{\text{\tiny EV1}}_{\text{\tiny EV2}}$ (e) and the relative local errors  $  \delta^{\text{\tiny PM}}_{\text{\tiny EV1}}$ (b), $  \delta^{\text{\tiny PM}}_{\text{\tiny EV2}}$ (d)  $\delta^{\text{\tiny EV1}}_{\text{\tiny EV2}}$ (f), evaluated using Eqs.\eqref{eq:error} from the density profiles presented in Fig.~\ref{fig:interface}, at 5 values of the temperature $T$.
\label{fig:error}}
\end{figure}

The resulting density profiles of the stationary interface in the final state are shown in  Fig. \ref{fig:interface}. In this figure, as well as in Fig.~\ref{fig:error}, we took advantage of both the symmetry of the computing domain and the periodic boundary conditions.  Consequently,  the plots were restricted to a limited interval on the positive $x$ axis, centered in the point $x_0$. For each value of $T$, the black dotted lines in Fig.~\ref{fig:interface} indicate the corresponding values of both the liquid and the vapor density, as calculated according to the Maxwell construction.  Far from the interface, where the density gradient vanishes, the PM, EV1, and EV2 density values are quite identical and well-superposed to the density values retrieved using the Maxwell construction, as observed in Fig.~\ref{fig:phase_diagram}.
These overlapping values agree very well with the theoretical predictions of the Maxwell construction. The agreement is illustrated also in Fig.~\ref{fig:phase_diagram}, which shows the phase diagram traced with the Maxwell construction, {\color{blue}as well as the PM, EV1 and EV2 values of the liquid and the vapor densities, obtained for the same values of $T$ as in Fig. \ref{fig:interface}.}

{\color{blue} {For all the temperature values considered in Fig. \ref{fig:interface}, one can easily observe the existing differences between the PM, the EV1 and the EV2 density profiles in the interface region. More precisely, the EV1 profiles seem to be closer to the PM profiles than the corresponding EV2 profiles.
The discrepancies between the PM and the EV1 profiles are attributed to the simplified Enskog collision operator \eqref{eq:joplusj1} employed to get the EV1 profiles, as the force field evaluation is identical in both the PM and the EV1 cases. At the same time, for all the five temperatures in Fig. \ref{fig:interface}, one observes that the discrepancies between the PM and the EV2 profiles are always larger with respect to the PM - EV1 discrepancies. This is not a surprise because, although both the EV1 and the EV2 profiles are plagued by the errors introduced by the simplified Enskog collision operator, the EV2 profiles are plagued also by the additional errors generated by the low order expansion of $\bm{\mathcal{F}}[n]$, used to evaluate the mean force field according to \eqref{eq:F2}. In general, both the PM - EV1 and the PM - EV2 discrepancies always increase when the temperature $T$ goes further below the critical value $T_c$.}}

For a quantitative evaluation of the differences between {\color{blue}the PM - EV1, PM - EV2 and  EV1 - EV2} density profile pairs  across the interface, in Figures \ref{fig:error} (a-f) we plot  the local error $\Delta^a_b(x)$ and the relative local error $\delta^a_b(x)$, defined as:
{\color{blue}
\begin{equation}\label{eq:error}
 \Delta^a_b(x) = \rho_a(x)-\rho_b(x);
 \quad \delta^a_b(x)=\frac{\Delta^a_b(x)}{\,\rho_a(x)\,}.
\end{equation}
}
{\color{blue}
The local error and the local relative errors introduced by the simplified Enskog collision term \eqref{eq:joplusj1} are $\Delta^{\text{\tiny PM}}_{\text{\tiny EV1}}(x)$ and $\delta^{\text{\tiny PM}}_{\text{\tiny EV1}}(x)$. These errors are shown in Fig.\ref{fig:error} (a) and (b) for the same temperature values as in Fig. \ref{fig:interface}. As seen in these figures, the largest values of the relative deviation are located in the interface region next to the vapour phase and increase substantially  (up to $60\%$) when decreasing the EV fluid temperature to $T=0.55$. Inspection of the $\Delta^{\text{\tiny PM}}_{\text{\tiny EV2}}(x)$ and the $\delta^{\text{\tiny PM}}_{\text{\tiny EV2}}(x)$ errors in Fig.\ref{fig:error} (c) and (d)
let us clearly know that the EV2 density profiles are subjected to larger errors than the EV1 profiles
for all the five values of the temperatures. As mentioned previously, this happens because of the expansion \eqref{eq:F2} used during the evaluation of the self-consistent force field, which introduces additional errors in the EV2 profiles. The local error and the local relative error introduced by the approximate expression \eqref{eq:F2} of the self-consistent (Vlasov) force is quantified in Fig.\ref{fig:error} (e) and fb) by $\Delta^{\text{\tiny EV1}}_{\text{\tiny EV2}}(x)$  and $\delta^{\text{\tiny EV1}}_{\text{\tiny EV2}}(x)$.  The local relative error in the latter figure exhibits larger values near the vapor phase than in the vicinity of the liquid phase. For example, at $T=0.55$, one can observe that the EV2 density is overestimating the corresponding EV1 value by more than $45\%$.} Following these results, we expect to see similar behavior in the numerical experiments presented in the next subsections.

\subsection{Phase separation dynamics in an Enskog-Vlasov fluid}\label{sec:phase_separation}

Let us consider an Enskog-Vlasov fluid whose state is periodic along the $x$ direction and homogeneous along the $y$ and $z$ directions. In the initial state ($t = 0$), the fluid density is subjected to a sinusoidal perturbation of wavelength $L=100\sigma$ and  amplitude $\omega$:
\begin{equation}\label{eq:rho_init_takata}
\rho_{i}(x) \equiv \rho(x,t=0) = A \left[1 + \omega \cos\left( \frac{ 2\pi x} {L}\right)\right]
\end{equation}
The simulation is conducted in the interval $[-L/2,\, +L/2]$ of the $x$ axis, with periodic boundary conditions.
For simplicity, the initial value of the fluid temperature $T_{i}$ is supposed to be uniform in the whole computational domain.  In Eq.~\eqref{eq:rho_init_takata}  above, we set  $A=(\rho_\ell+\rho_v)/2$, where $\rho_\ell$ and $\rho_v$ are the values of the liquid and the vapor density, evaluated using the Maxwell construction at temperature $T_{i}$. Because of the symmetry with respect to the origin of the $x$ axis, the plots in all figures belonging to this Subsection were restricted to the positive semi-axis.

In order to avoid the occurrence of spontaneous spinodal decomposition in the PM  simulations due to inherent noise, in all simulations conducted in this Subsection we have chosen the following set of initial temperatures $T_i$ and amplitudes $\omega$: $(T_i,\omega)\in\{(0.7,0.4);(0.65,0.55);(0.6,0.65);(0.55,0.7)\}$. For each $T_i$, the associated value of $\omega$ was chosen in order to ensure that both the maximum and the minimum values of $\rho_{i}(x)$, calculated according to Eq.~\eqref{eq:rho_init_takata}, lie outside the spinodal interval (i.e., the interval between the two spinodal curves obtained by requiring that the pressure gradient with respect to density to vanish at constant temperature).

\begin{figure}
\begin{tabular}{ccc}
\includegraphics[width=0.33\linewidth]{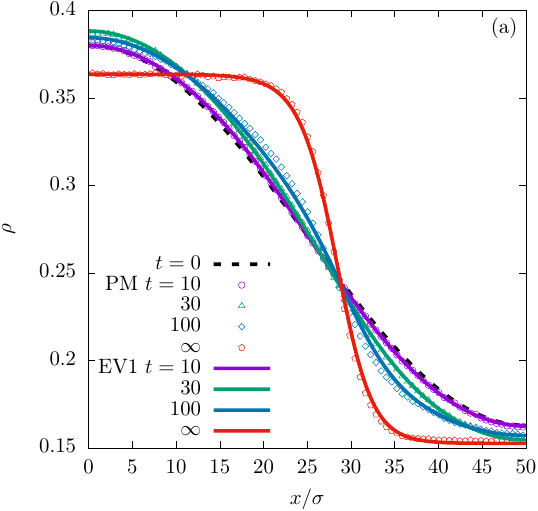}&
\includegraphics[width=0.33\linewidth]{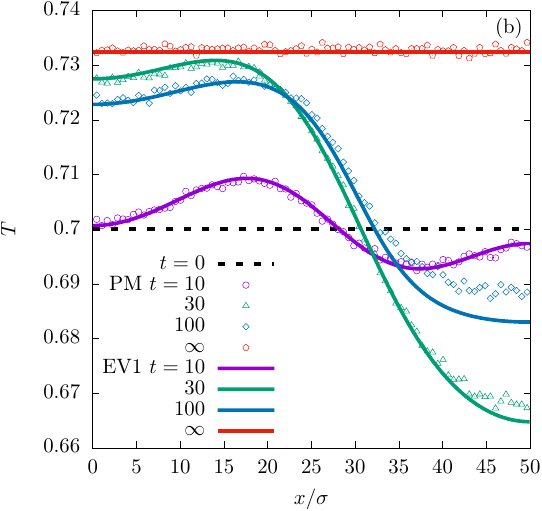}&
\includegraphics[width=0.33\linewidth]{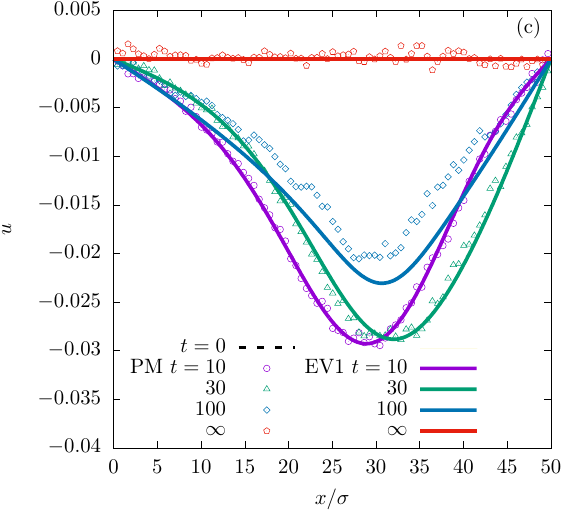}\\
\includegraphics[width=0.33\linewidth]{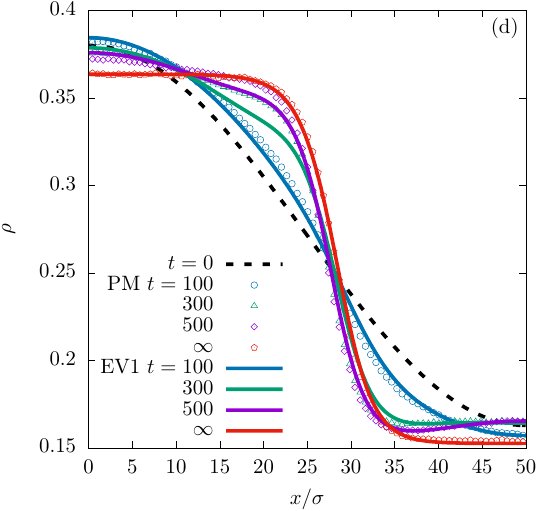}&
\includegraphics[width=0.33\linewidth]{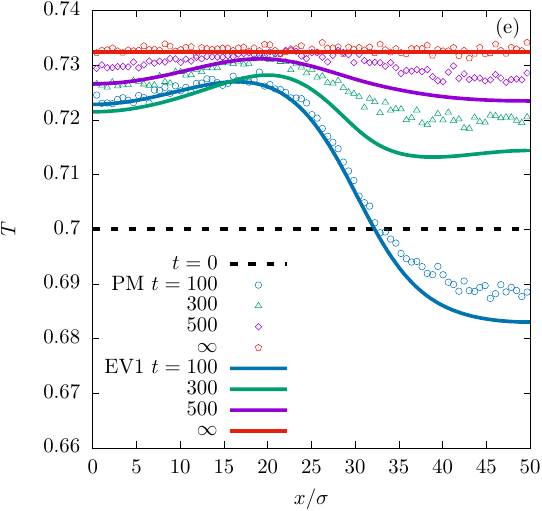}&
\includegraphics[width=0.33\linewidth]{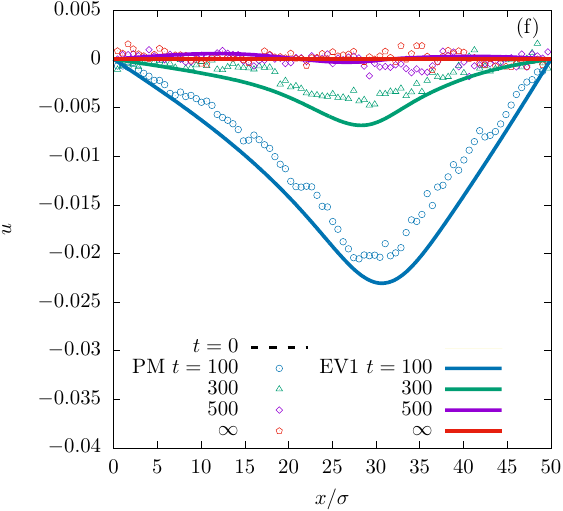}
\end{tabular}
\caption{
Phase separation with $(T_i, \omega) = (0.7, 0.4)$: comparison between PM and EV1 profiles of density  (a,d), (temperature b,e) , and  macroscopic velocity (c,f) at various time instances $t\in\{0, 10, 30, 100, 300, 500, \infty \}$.  The $\infty$ symbol refers to the quasi-stationary state ($t=5000$). For better readability, the profiles at $t\le 100$  and  $t \ge 100$ are shown in the first and the second row, respectively.
}\label{fig:takata_vartime}
\end{figure}

\begin{figure}
\begin{tabular}{ccc}
\includegraphics[width=0.33\linewidth]{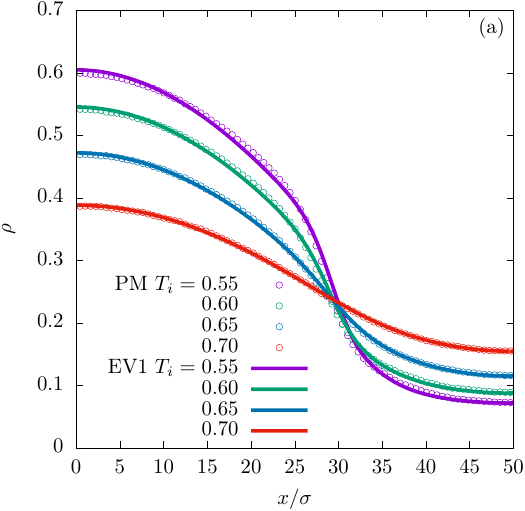}&
\includegraphics[width=0.33\linewidth]{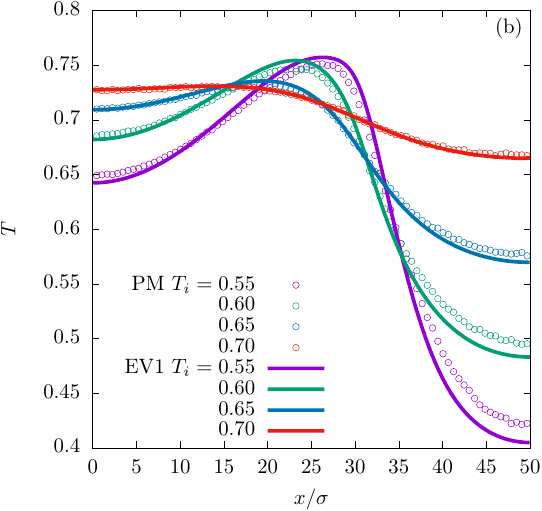}&
\includegraphics[width=0.33\linewidth]{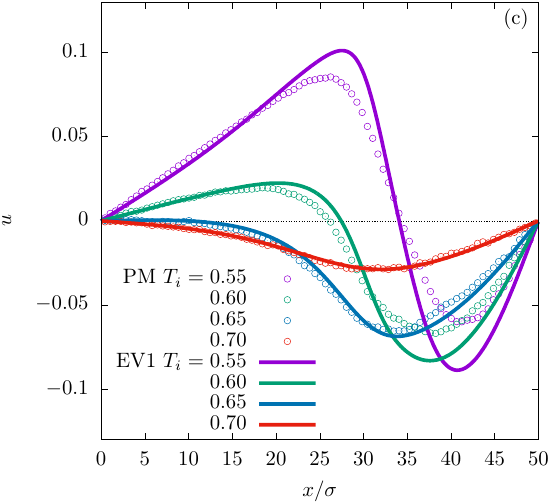}
\end{tabular}
\caption{
Phase separation at $t = 30$: comparison between PM and EV1 profiles of  density (a), temperature (b) e and  macroscopic velocity (c) for four values of the initial temperature $T_i$.
}\label{fig:takata_vartemp}
\end{figure}

The PM and the EV1 simulation results obtained for the pair of initial conditions $(T_i, \omega) = (0.7, 0.4)$ are compared in Fig.~\ref{fig:takata_vartime}, where we deliberately choose an initial temperature close to the critical one. In this figure, the density, temperature, and macroscopic velocity profiles are compared at six time instances $t\in\{10,30,100,300,500,\infty\}$, where the symbol $\infty$ refers to the stationary state.  Reasonable agreement of all PM and EV1  profiles is observed in Fig.~\ref{fig:takata_vartime} for all values of $t$. During the phase separation, oscillations of both the temperature and the macroscopic velocity profiles are observed until the stationary state is reached. Small discrepancies between the PM and the EV1 profiles of both the temperature and the macroscopic velocity are noticed especially in the vapor phase ($x/\sigma > 25$), but these discrepancies rarely exceed $5\%$.  Since the force field evaluation is identical in both the PM and the EV1 simulations, these discrepancies may be attributed to the simplified Enskog collision operator.
Despite these deviations, the EV1 stationary profiles ($t = \infty$) of the density, temperature, and macroscopic velocity are well superposed to the PM profiles.  After the phase transition ends and the stationary state is reached, we note that the temperature profile is a constant whose value is always higher than the initial value $T_i$. As explained in \cite{takata2021kinetic}, during the phase separation process, a part of the internal energy which is due to the attractive molecular interaction is released and spent on the temperature rise.

In Fig.~\ref{fig:takata_vartemp}, we compare the density, the temperature, and the macroscopic velocity profiles resulted in PM and EV1 simulations at time $t=30$, for 4 values of the initial temperature $T_i$.
The evolution of these profiles at various values of the initial temperature is similar to the case shown in Fig.~\ref{fig:takata_vartime}. In general, the discrepancies between the PM and EV1 profiles of the temperature and the macroscopic velocity become larger when the value $T_i$ of the initial temperature descends farther below the critical one $T_c$. {\color{blue}This is attributed to the density of the liquid, which increases as temperature decreases. As noted in previous studies \cite{B23,BS24,B24}, a higher density introduces significant errors due to the simplifications in the Enskog collision operator.}

In Fig. \ref{fig:takata_varEV}, we compare the simulation results obtained using the two implementations of the self-consistent force field, EV1 and EV2, with the corresponding PM results. We choose the lowest and the highest temperatures already tested, namely $T_i\in\{0.55,0.7\}$, and plot the three usual profiles at $t=100$. The top row refers to the lowest temperature, while the three columns present the density, temperature, and macroscopic velocity, respectively. While there are some small differences between the two implementations (EV1 and EV2), they match very well throughout the flow domain, both for small and large temperatures. {\color{blue} {We attribute this rather unexpected behavior to the sinusoidal profile of the fluid density in the initial state, where the density gradients are relatively small
and thus, the additional errors introduced by the self-consistent (Vlasov) force \eqref{eq:F2} in the EV2 profiles become negligible with respect to the errors introduced by the simplified collision term in both the EV1 and the EV2 simulations. Due to the lower computational time demanded by the EV2 procedure (see Appendix \ref{app:runtime}), its use becomes advantageous when simulating dynamic problems involving small gradients.}}

\begin{figure}
\begin{tabular}{ccc}
\includegraphics[width=0.33\linewidth]{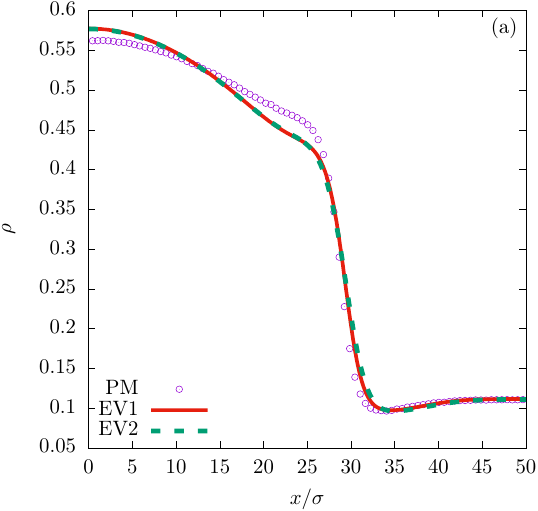}&
\includegraphics[width=0.33\linewidth]{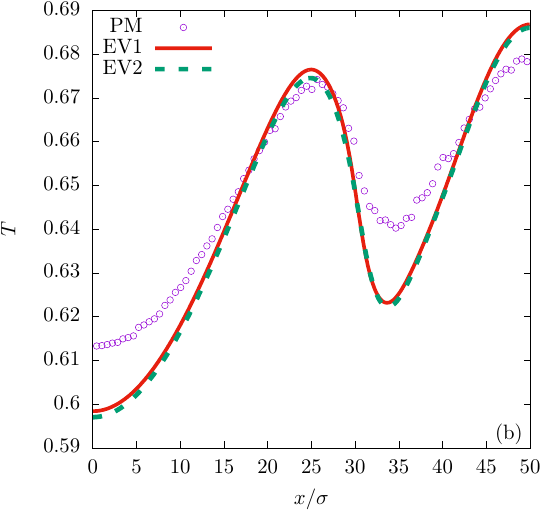}&
\includegraphics[width=0.33\linewidth]{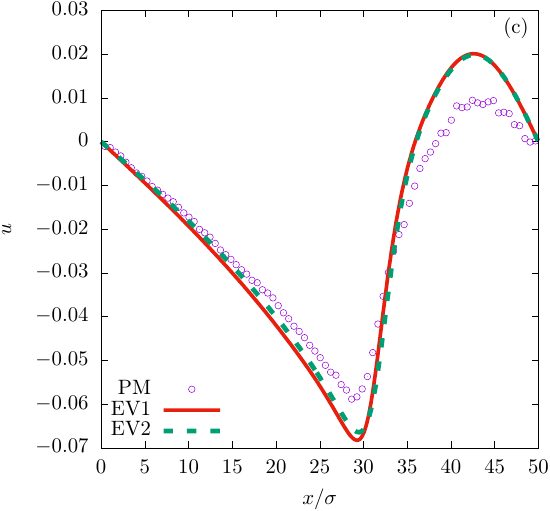}\\
\includegraphics[width=0.33\linewidth]{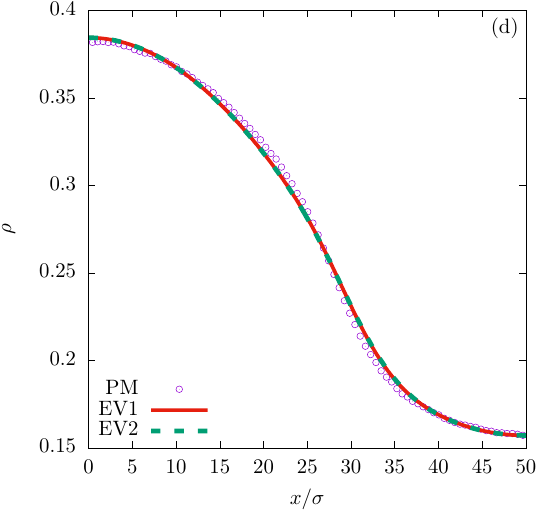}&
\includegraphics[width=0.33\linewidth]{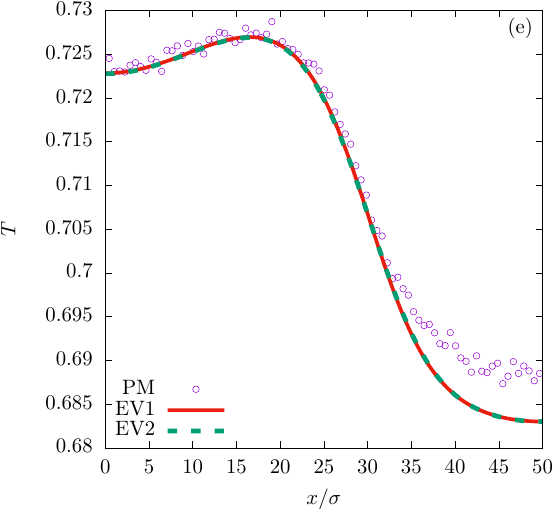}&
\includegraphics[width=0.33\linewidth]{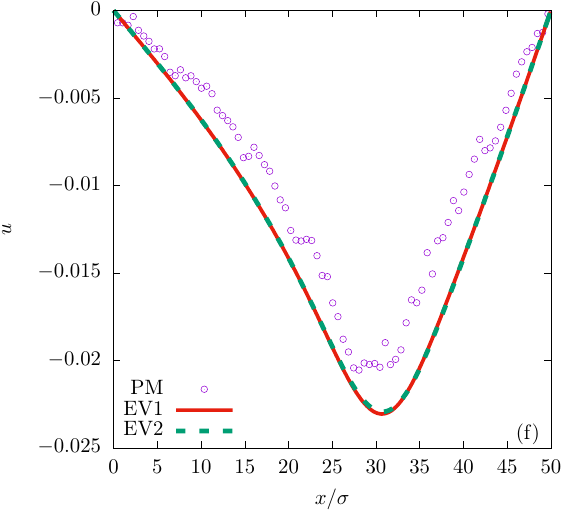}
\end{tabular}
\caption{
Phase separation: density, temperature, and macroscopic velocity profiles calculated at $t = 100$ using the PM, EV1, and EV2 methods, for $T_i = 0.55$ (a-c) and $T_i = 0.7$ (d-f).
} \label{fig:takata_varEV}
\end{figure}

\subsection{Evaporation}\label{sec:evaporation}

\begin{figure}
\begin{tabular}{ccc}
\includegraphics[width=0.32\linewidth]{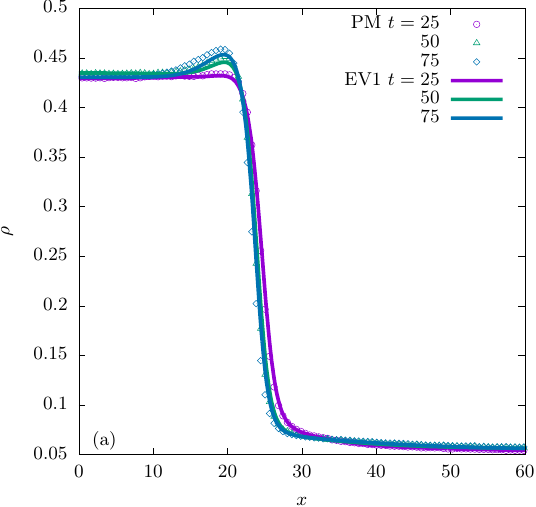}
\includegraphics[width=0.32\linewidth]{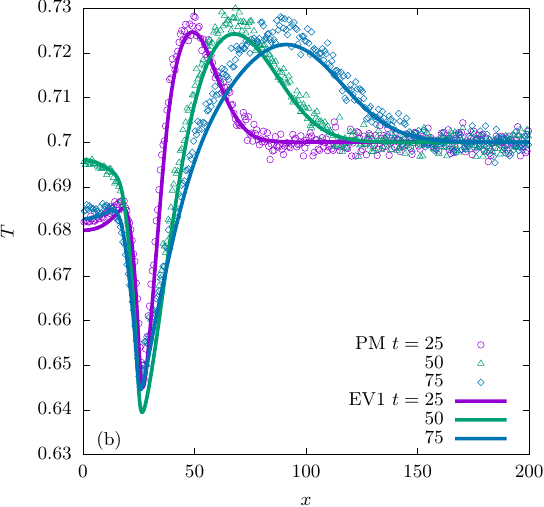}
\includegraphics[width=0.32\linewidth]{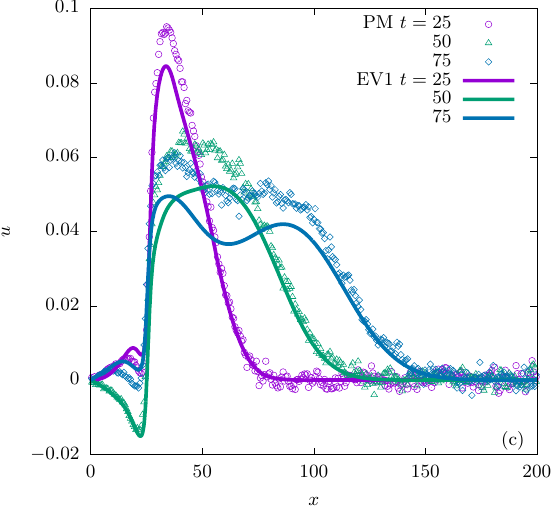}
\end{tabular}
\caption{Evaporation into metastable vapor: evolution of the density, temperature, and macroscopic velocity profiles calculated at 3 values of $t$ using both the PM and the EV1 methods (initial value of the fluid temperature $T_i = 0.70$).}\label{fig:evap_vartime}
\end{figure}

\begin{figure}
\begin{tabular}{ccc}
\includegraphics[width=0.33\linewidth]{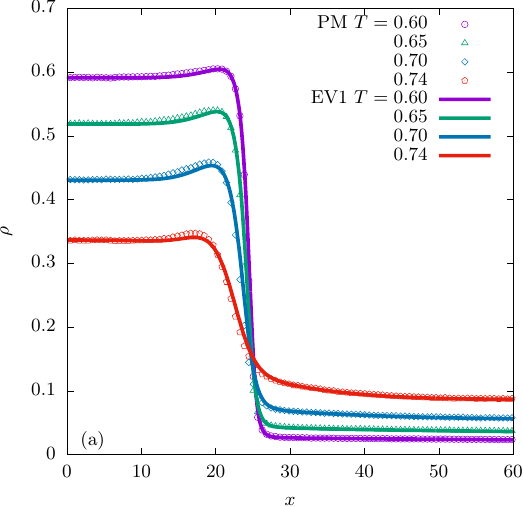}&
\includegraphics[width=0.33\linewidth]{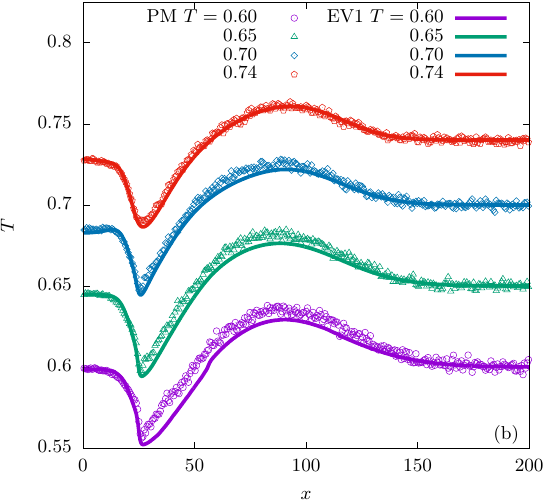}&
\includegraphics[width=0.33\linewidth]{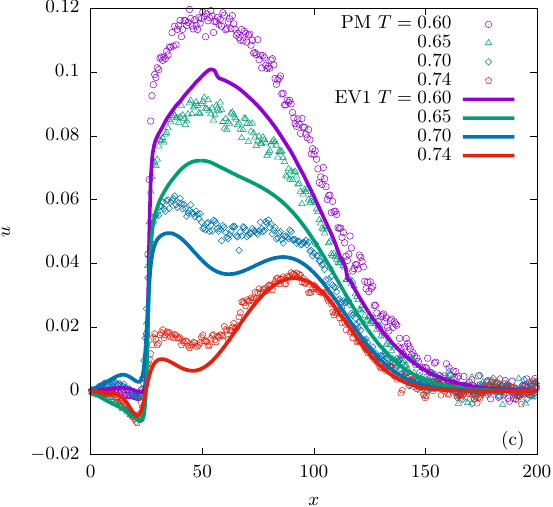}
\end{tabular}
\caption{Evaporation into metastable vapor: fluid density, temperature, and macroscopic velocity profiles calculated at $t = 75$ for 4 values of the initial temperature $T_{i}$ using the PM and the  EV1 methods.}\label{fig:evap_varT}
\end{figure}

The evaporation of a liquid slab into its metastable vapor is studied in this section. Initially, the liquid slab is centered at the origin of the real axis, with symmetric smooth interfaces located at $x_0=\pm 25\sigma$ and periodic boundary conditions. Let $\rho_\ell$ and $\rho_v$ be the liquid and the vapor density values calculated according to the Maxwell construction at the initial temperature $T_i$ of the fluid system. At $t=0$, the liquid density value in the slab is set to $\rho_\ell$, while the vapor density value outside the slab is set to $\rho_v / 2$. In all simulations reported in this subsection, the computational domain is located within the interval $[\, -200\sigma,\,200\sigma\,]$ of the real axis. Since the computational domain is symmetric with respect to the origin of the $x$ axis, the corresponding plots were restricted to a smaller part of the positive semi-axis. At the beginning of each simulation, the left and the right  liquid-vapor interface profiles are set according to:
\begin{equation}\label{eq:tan}
 \rho_{i}(x)=\rho_\ell + \frac{\rho_v - 2\rho_\ell}{ 4} \left[1 + \tanh\left(\frac{|x| - x_0 }{ \xi}\right)\right];
\end{equation}
where $\xi=\sqrt{ \frac{3}{ 4 \pi \sqrt{3}  (T_c - T_i)}}$ and $x_0=25 \sigma$.

The values $T_i$ of the initial temperature used in the simulations reported in this subsection are $T_i\in\{0.6,0.65,0.7,0.74\}$. Unlike the previous section, the lowest temperature employed in this section is $T_i=0.6$, because of the unstable behaviour arising due to the sharp gradient involved. The time evolution of the system at  $T_i=0.7$ is presented in Fig.~\ref{fig:evap_vartime}. In these plots, we compare the density, the temperature, and the macroscopic velocity profiles obtained at $t\in\{25,50,75\}$ with both the PM and the EV1 methods. The time instances were chosen such that the periodic boundary conditions would not interfere with the evolution of the fluid.
Since the value of the initial temperature $T_i$ is not too far from the critical temperature $T_c$, one may observe that the EV1 results follow closely the PM results throughout the time interval chosen, which stretches spatially eight times the initial slab length $x_0$. Reasonable accuracy is observed for all quantities.

\begin{figure}
\begin{tabular}{ccc}
\includegraphics[width=0.33\linewidth]{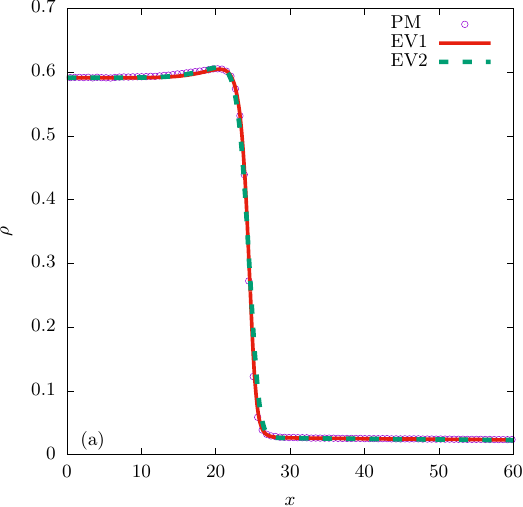}&
\includegraphics[width=0.33\linewidth]{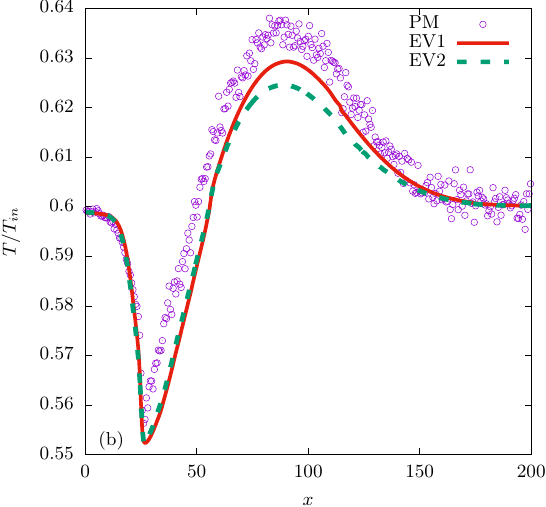}&
\includegraphics[width=0.33\linewidth]{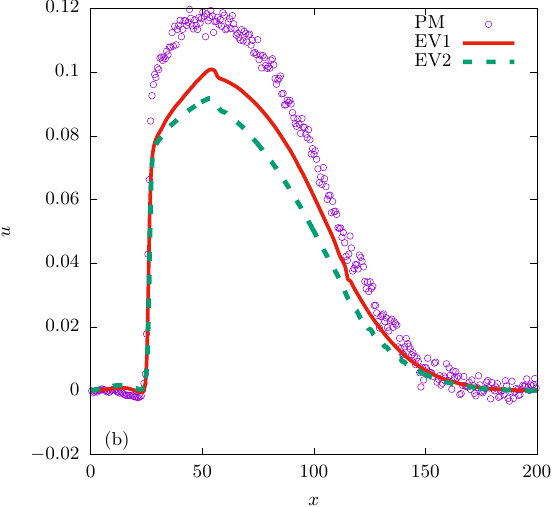}\\
\includegraphics[width=0.33\linewidth]{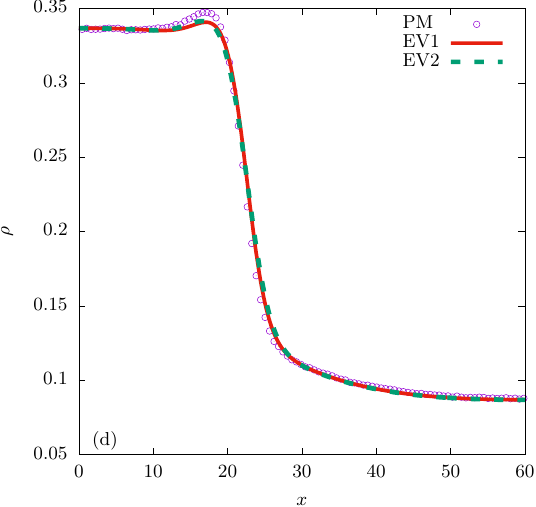}&
\includegraphics[width=0.33\linewidth]{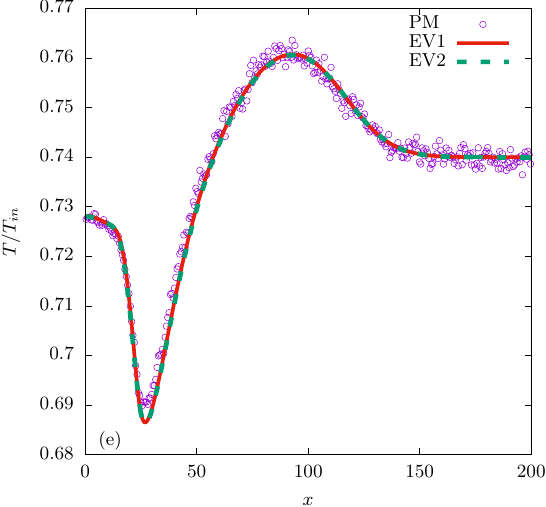}&
\includegraphics[width=0.33\linewidth]{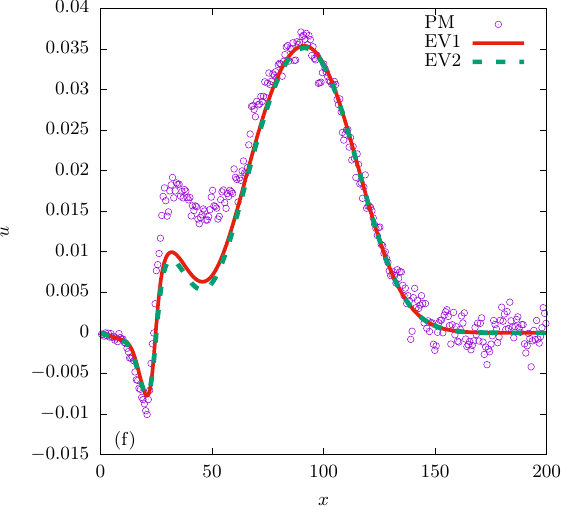}
\end{tabular}
\caption{ Evaporation into metastable vapor: fluid density, temperature, and velocity profiles calculated at time $t=75$ using  the PM, EV1, and EV2 methods/implementations, for $T_i = 0.6$ (a-c) and $T_i = 0.74$ (d-f).}\label{fig:evap_varEV}
\end{figure}

Next, we investigate the accuracy of the EV1 results obtained at $t=75$ for various values of the initial temperature $T_i$. The density, the temperature, and the macroscopic velocity profiles obtained with
both the PM and the EV1 methods are shown in Figs.\ref{fig:evap_varT} (a-c). As the temperature decreases, we observe that the deviations between the  EV1 and the PM profiles of the temperature and the macroscopic velocity become more pronounced, although in good qualitative agreement.

Finally, we compare the two implementations of the self-consistent force field, EV1 and EV2. We choose the lowest and the highest temperatures already tested, namely $T_i\in\{0.6,0.74\}$, and plot the three usual profiles at $t=75$. The results are gathered in Fig.~\ref{fig:evap_varEV}, where the top row refers to the lowest temperature. The three columns contain the density, temperature, and macroscopic velocity, respectively. While at high temperatures there is no difference between the two implementations, as expected due to the small gradient appearing at the interface, at the lowest temperature one can observe that the EV2 results deviate from the PM results further away than the EV1 results, due to the approximation in evaluating the molecular attraction{\color{blue}{, i.e. the Vlasov force term \eqref{eq:force}}}. On the other hand, due to its much lower computational demand (see Appendix \ref{app:runtime}), the EV2 implementation still proves to be a convenient alternative to EV1, especially for exploratory (preliminary) purposes and/or large-scale simulations.

\section{CONCLUSION}\label{sec4}

In this work, an Enskog-Vlasov finite-difference Lattice Boltzmann (EV-FDLB) model with variable temperature was developed from an existing Enskog FDLB model involving the simplified Enskog collision operator. Two procedures (denoted EV1 and EV2) for evaluating the self-consistent force field integral were considered: a first one that evaluates the integral directly and a second one that approximates the integral by assuming that the density field is sufficiently smooth anywhere, including the phase interface. The simulation results were further compared with a Particle Method (PM) which solves the full Enskog collision integral and evaluates the mean-force field using the first procedure.

We first compared the isothermal stationary interface profiles obtained using {\color{blue} the PM, as well as the }EV1 and EV2 procedures. {\color{blue} The fluid density values retrieved using the PM, as well as the EV1 and EV2 procedures, overlap very well in both the liquid and the vapour phase, i.e., outside the liquid-vapor interface, where the density gradients vanish. In the pure phases, all these values are in excellent agreement with the phase diagram evaluated using the Maxwell construction. As} the temperature decreases and the interface becomes sharper, {\color{blue} both the EV1 and} the EV2 results start diverging from the {\color{blue} PM} results, especially in the vapor side of the interface area. {\color{blue} More precisely, the discrepancies between the EV2 and the PM profiles are larger than the corresponding EV1 and PM profiles, for the five values of the temperature considered.  Although the simplified Enskog collision operator is used in both the EV1 and the EV2 simulations, the EV2 simulations involve the series expansion of the mean field (Vlasov) force  \eqref{eq:F2}, which introduces a supplementary error with respect to the EV1 results. This supplementary error was found to increase when the fluid temperature decreased. For example, a local relative error of around $45\%$ is observed at $T=0.55$ in the interface region close to the vapor phase between the EV1 and EV2 density profiles.  Note that, because of the smaller value of the vapor density, this relative error value is much larger than the corresponding error close to the liquid phase.}

Since the EV2 procedure was found to be less expensive than the EV1 procedure, the main goal of our work was to assess the accuracy of the EV2 simulation results with respect to the corresponding EV1 and PM results. For this purpose, two dynamic problems involving liquid-vapor phase separation and variable temperature were considered in this paper.

The first thermal problem studied was the one-dimensional phase separation dynamics triggered by a sinusoidal density profile. We compared the EV1 and EV2 to the PM and we observed that the two procedures give similar results at temperatures close to the critical one, but they diverge from the PM results as the initial temperature decreases. This is attributed to the liquid density, which has increasing values as temperature decreases. As already noticed in previous works\cite{B23,BS24,B24}, a larger value of density brings significant errors due to the simplified Enskog collision operator. Overall, reasonable agreement is achieved for the density, temperature, and macroscopic velocity.

In the second thermal problem, we looked at the initial stages of the evaporation of a liquid slab into metastable vapor. At high values of the initial temperatures, both implementations show similar results, which is expected due to the small gradient at the interface. However, at low temperatures, the EV2 results deviate more from the PM results compared to the EV1 results. This is attributed to the approximation used in evaluating molecular attraction in EV2. Despite this, the EV2 implementation stands out as a strong alternative to EV1, particularly due to its significantly lower computational demand, at least for the temperatures tested in this study.

{\color{blue} In conclusion, we hope that the present Enskog-Vlasov model demonstrated its capability to handle thermal liquid-vapor flows using a lattice Boltzmann approach. Moving forward, an important goal will be the introduction of appropriate boundary conditions and the extension to two- or three-dimensional multiphase systems with variable temperature.}

\begin{acknowledgments}
This work was supported through a grant from the Ministry of Research, Innovation and Digitization, CNCS-UEFISCDI Project No. PN-III-P1-1.1-PD-2021-0216, within PNCDI III.
\end{acknowledgments}

\appendix

\section{Particle Method of Solution}\label{app:PM}

In this work, the EV equation is numerically solved using an extension of the original Direct Simulation Monte-Carlo (DSMC) scheme tailored for dense fluids \cite{F97b}. A detailed description of the numerical scheme and its computational complexity is provided in Ref.~\cite{FBG19}.

For the EV simulations, the core framework of the DSMC scheme used to solve the Boltzmann equation is maintained, with modifications in the collision algorithm due to the nonlocal nature of the Enskog collision operator. The distribution function is represented by $N$ computational particles:
\begin{equation}
f(\bm{x},\bm{p},t)= \frac{1}{m} \sum_{i=1}^{N} \delta{\left(\bm{x}-\bm{x}_i(t)\right)} \delta(\bm{p}-\bm{p}_i(t)),
\end{equation}
where $\bm{r}_i$ and $\bm{p}_i$ are the position and the momentum of the $i$th particle at time $t$, respectively.

The distribution function is updated using a fractional-step method based on time-splitting the evolution operator into two sub-steps: free streaming and collision. In the first stage, particle collisions are neglected, and the distribution function is advanced from $t$ to $t + \Delta t$ by solving the equation:
\begin{equation}
\label{eq:stage_I}
    \frac{\partial f}{\partial t} +\frac{\bm{p}}{m}\cdot\nabla_{\bm{x}}f + \bm{\mathcal{F}}_1[n] \cdot \nabla_{\bm{p}} f = 0,
\end{equation}
which translates into updating the positions and velocities of the computational particles according to:
\begin{subequations}
\begin{align}
 \bm{r}_i(t + \Delta t) &= \bm{r}_i(t) +\frac{\bm{p}_i}{m}\Delta t + \frac{\bm{\mathcal{F}}_1[n(t)]}{m} \frac{(\Delta t)^2}{2},\\
\bm{p}_i(t + \Delta t) &= \bm{p}_i(t) + \frac{\bm{\mathcal{F}}_1[n(t)]}{m} \Delta t.
 \end{align}
\end{subequations}
In the second stage, short-range hard-sphere interactions are considered, and the updating rule is given by:
\begin{equation}
 f(\bm{r}, \bm{p}, t + \Delta t) = \tilde{f}(\bm{r}, \bm{p}, t + \Delta t) + J_E[\tilde{f}] \Delta t.
\end{equation}
During this stage, particle positions $\bm{r}_i$ remain unchanged while their momenta $\bm{p}_i$ are modified according to stochastic rules, which essentially correspond to the Monte Carlo evaluation of the collision integral given by Eq.~\eqref{eq:enskog}.
An average number of 10000 particles per cell were used, a cell size of $\Delta x=\sigma/10$, and the time step was set to $\Delta t=10^{-3}$. {\color{blue} These values were chosen following a convergence test on all variables.}
The macroscopic quantities are obtained by time-averaging the particles' microscopic states (every 1000 iterations) as well as state-averaging by running 20 simulations with random seeds.

\section{Runtime comparison}\label{app:runtime}

\begin{table}
	\begin{center}
		\begin{tabular*}{\columnwidth}{@{\extracolsep{\stretch{1}}}*{7}{l|cc|cc}@{}}
			\toprule
			   & Sec.\ref{sec:phase_separation}  & &  Sec.\ref{sec:evaporation}  \\ \hline
           & $T=0.55$ & $T=0.70$ & $T=0.60$ & $T=0.74$  \\ \hline \hline
			  $t_{\text{\tiny PM}}$(for 20 runs) &  $8.2\times10^6$s  & $6\times10^6$s &  $10^6$s &  $8.4\times10^5$s  \\ \hline
			 $t_{\text{\tiny EV1}}$  &  $1.8\times10^4$   & $1.8\times10^4$s & $3.4\times10^3$s &  $3.4\times10^3$s    \\ \hline
       $t_{\text{\tiny EV2}}$  & $ 6\times10^3$s  & $ 6\times10^3$s & $ 9\times10^2$s &  $ 9\times10^2$s \\ \hline\hline
       $t_{\text{\tiny PM}}/t_{\text{\tiny EV1}}$   &  455  & 333 & 295  &  240 \\ \hline
        $t_{\text{\tiny PM}}/t_{\text{\tiny EV2}}$   &  1365  & 1000 & 1110  &  930 \\ \hline
			 \hline
		\end{tabular*}
	\end{center}
	\caption{Typical runtimes and runtime ratios for the simulations presented in Secs. \ref{sec:phase_separation} and \ref{sec:evaporation}. }
	\label{tab:comp_time}
\end{table}

The typical runtimes for the simulation conducted with the PM, EV1 and EV2 methods, as well as the runtime ratios, are presented in Table \ref{tab:comp_time} for the setups in Secs. \ref{sec:phase_separation} and \ref{sec:evaporation}. The runtimes for EV1 and EV2 are independent of the initial temperature. These times were recorded on a single core of an Intel(R) Xeon(R) Gold 6330 CPU running at 2.0GHz.

\bibliography{bibliography.bib}

\providecommand{\noopsort}[1]{}\providecommand{\singleletter}[1]{#1}%
\begin{thebibliography}{65}%
\makeatletter
\providecommand \@ifxundefined [1]{%
 \@ifx{#1\undefined}
}%
\providecommand \@ifnum [1]{%
 \ifnum #1\expandafter \@firstoftwo
 \else \expandafter \@secondoftwo
 \fi
}%
\providecommand \@ifx [1]{%
 \ifx #1\expandafter \@firstoftwo
 \else \expandafter \@secondoftwo
 \fi
}%
\providecommand \natexlab [1]{#1}%
\providecommand \enquote  [1]{``#1''}%
\providecommand \bibnamefont  [1]{#1}%
\providecommand \bibfnamefont [1]{#1}%
\providecommand \citenamefont [1]{#1}%
\providecommand \href@noop [0]{\@secondoftwo}%
\providecommand \href [0]{\begingroup \@sanitize@url \@href}%
\providecommand \@href[1]{\@@startlink{#1}\@@href}%
\providecommand \@@href[1]{\endgroup#1\@@endlink}%
\providecommand \@sanitize@url [0]{\catcode `\\12\catcode `\$12\catcode
  `\&12\catcode `\#12\catcode `\^12\catcode `\_12\catcode `\%12\relax}%
\providecommand \@@startlink[1]{}%
\providecommand \@@endlink[0]{}%
\providecommand \url  [0]{\begingroup\@sanitize@url \@url }%
\providecommand \@url [1]{\endgroup\@href {#1}{\urlprefix }}%
\providecommand \urlprefix  [0]{URL }%
\providecommand \Eprint [0]{\href }%
\providecommand \doibase [0]{https://doi.org/}%
\providecommand \selectlanguage [0]{\@gobble}%
\providecommand \bibinfo  [0]{\@secondoftwo}%
\providecommand \bibfield  [0]{\@secondoftwo}%
\providecommand \translation [1]{[#1]}%
\providecommand \BibitemOpen [0]{}%
\providecommand \bibitemStop [0]{}%
\providecommand \bibitemNoStop [0]{.\EOS\space}%
\providecommand \EOS [0]{\spacefactor3000\relax}%
\providecommand \BibitemShut  [1]{\csname bibitem#1\endcsname}%
\let\auto@bib@innerbib\@empty
\bibitem [{\citenamefont {Chapman}\ and\ \citenamefont
  {Cowling}(1970)}]{cowling70}%
  \BibitemOpen
  \bibfield  {author} {\bibinfo {author} {\bibfnamefont {S.}~\bibnamefont
  {Chapman}}\ and\ \bibinfo {author} {\bibfnamefont {T.~G.}\ \bibnamefont
  {Cowling}},\ }\href@noop {} {\emph {\bibinfo {title} {The Mathematical Theory
  of Non-uniform Gases: An Account of the Kinetic Theory of Viscosity, Thermal
  Conduction and Diffusion in Gases.}}}\ (\bibinfo  {publisher} {Cambridge
  University Press},\ \bibinfo {year} {1970})\BibitemShut {NoStop}%
\bibitem [{\citenamefont {Ferziger}\ and\ \citenamefont {Kaper}(1972)}]{FK72}%
  \BibitemOpen
  \bibfield  {author} {\bibinfo {author} {\bibfnamefont {J.}~\bibnamefont
  {Ferziger}}\ and\ \bibinfo {author} {\bibfnamefont {H.}~\bibnamefont
  {Kaper}},\ }\href@noop {} {\emph {\bibinfo {title} {Mathematical Theory of
  Transport Processes in Gases.}}}\ (\bibinfo  {publisher} {North-Holland
  Publishing Company, Amsterdam, London},\ \bibinfo {year} {1972})\BibitemShut
  {NoStop}%
\bibitem [{\citenamefont {Dorfman}\ \emph {et~al.}(2021)\citenamefont
  {Dorfman}, \citenamefont {van Beijeren},\ and\ \citenamefont
  {Kirkpatrick}}]{DBK2021}%
  \BibitemOpen
  \bibfield  {author} {\bibinfo {author} {\bibfnamefont {J.~R.}\ \bibnamefont
  {Dorfman}}, \bibinfo {author} {\bibfnamefont {H.}~\bibnamefont {van
  Beijeren}},\ and\ \bibinfo {author} {\bibfnamefont {T.}~\bibnamefont
  {Kirkpatrick}},\ }\href
  {https://doi.org/https://doi.org/10.1017/9781139025942} {\emph {\bibinfo
  {title} {{Contemporary Kinetic Theory of Matter}}}}\ (\bibinfo  {publisher}
  {Cambridge University Press},\ \bibinfo {year} {2021})\BibitemShut {NoStop}%
\bibitem [{\citenamefont {Bird}(1976)}]{B76}%
  \BibitemOpen
  \bibfield  {author} {\bibinfo {author} {\bibfnamefont {G.~A.}\ \bibnamefont
  {Bird}},\ }\href@noop {} {\emph {\bibinfo {title} {Molecular Gas Dynamics}}}\
  (\bibinfo  {publisher} {Oxford Univ. Press, Oxford, England, UK},\ \bibinfo
  {year} {1976})\BibitemShut {NoStop}%
\bibitem [{\citenamefont {Bird}(2013)}]{B13}%
  \BibitemOpen
  \bibfield  {author} {\bibinfo {author} {\bibfnamefont {G.~A.}\ \bibnamefont
  {Bird}},\ }\href@noop {} {\emph {\bibinfo {title} {The DSMC Method}}}\
  (\bibinfo  {publisher} {CreateSpace, US},\ \bibinfo {year}
  {2013})\BibitemShut {NoStop}%
\bibitem [{\citenamefont {Bird}(2025)}]{BDSMC}%
  \BibitemOpen
  \bibfield  {author} {\bibinfo {author} {\bibfnamefont {G.~A.}\ \bibnamefont
  {Bird}},\ }\href@noop {} {\emph {\bibinfo {title} {DSMC}}}\ (\bibinfo
  {publisher} {Oxford Univ. Press, Oxford, England, UK},\ \bibinfo {year}
  {2025})\BibitemShut {NoStop}%
\bibitem [{\citenamefont {Akhlaghi}\ \emph {et~al.}(2023)\citenamefont
  {Akhlaghi}, \citenamefont {Roohi},\ and\ \citenamefont {Stefanov}}]{ARS23}%
  \BibitemOpen
  \bibfield  {author} {\bibinfo {author} {\bibfnamefont {H.}~\bibnamefont
  {Akhlaghi}}, \bibinfo {author} {\bibfnamefont {E.}~\bibnamefont {Roohi}},\
  and\ \bibinfo {author} {\bibfnamefont {S.}~\bibnamefont {Stefanov}},\
  }\bibfield  {title} {\bibinfo {title} {A comprehensive review on micro- and
  nano-scale gas flow effects: {S}lip-jump phenomena, {K}nudsen paradox,
  thermally-driven flows, and {K}nudsen pumps},\ }\href
  {https://doi.org/10.1016/j.physrep.2022.10.004} {\bibfield  {journal}
  {\bibinfo  {journal} {Physics Reports}\ }\textbf {\bibinfo {volume} {997}},\
  \bibinfo {pages} {1} (\bibinfo {year} {2023})}\BibitemShut {NoStop}%
\bibitem [{\citenamefont {Frezzotti}(1997)}]{F97b}%
  \BibitemOpen
  \bibfield  {author} {\bibinfo {author} {\bibfnamefont {A.}~\bibnamefont
  {Frezzotti}},\ }\bibfield  {title} {\bibinfo {title} {{A particle scheme for
  the numerical solution of the Enskog equation}},\ }\href
  {https://doi.org/10.1063/1.869247} {\bibfield  {journal} {\bibinfo  {journal}
  {Physics of Fluids}\ }\textbf {\bibinfo {volume} {9}},\ \bibinfo {pages}
  {1329} (\bibinfo {year} {1997})}\BibitemShut {NoStop}%
\bibitem [{\citenamefont {De~Sobrino}(1967)}]{S67}%
  \BibitemOpen
  \bibfield  {author} {\bibinfo {author} {\bibfnamefont {L.}~\bibnamefont
  {De~Sobrino}},\ }\bibfield  {title} {\bibinfo {title} {On the kinetic theory
  of a van der {W}aals gas},\ }\href@noop {} {\bibfield  {journal} {\bibinfo
  {journal} {Can. J. Phys.}\ }\textbf {\bibinfo {volume} {45}},\ \bibinfo
  {pages} {363} (\bibinfo {year} {1967})}\BibitemShut {NoStop}%
\bibitem [{\citenamefont {Grmela}(1971)}]{G71}%
  \BibitemOpen
  \bibfield  {author} {\bibinfo {author} {\bibfnamefont {M.}~\bibnamefont
  {Grmela}},\ }\bibfield  {title} {\bibinfo {title} {Kinetic equation approach
  to phase transitions},\ }\href@noop {} {\bibfield  {journal} {\bibinfo
  {journal} {J. Stat. Phys.}\ }\textbf {\bibinfo {volume} {3}},\ \bibinfo
  {pages} {347} (\bibinfo {year} {1971})}\BibitemShut {NoStop}%
\bibitem [{\citenamefont {Karkheck}\ and\ \citenamefont {Stell}(1981)}]{KS81}%
  \BibitemOpen
  \bibfield  {author} {\bibinfo {author} {\bibfnamefont {J.}~\bibnamefont
  {Karkheck}}\ and\ \bibinfo {author} {\bibfnamefont {G.}~\bibnamefont
  {Stell}},\ }\bibfield  {title} {\bibinfo {title} {Kinetic mean-field
  theories},\ }\href@noop {} {\bibfield  {journal} {\bibinfo  {journal} {J.
  Chem. Phys.}\ }\textbf {\bibinfo {volume} {75}},\ \bibinfo {pages} {1475}
  (\bibinfo {year} {1981})}\BibitemShut {NoStop}%
\bibitem [{\citenamefont {He}\ and\ \citenamefont {Doolen}(2002)}]{HD02}%
  \BibitemOpen
  \bibfield  {author} {\bibinfo {author} {\bibfnamefont {X.}~\bibnamefont
  {He}}\ and\ \bibinfo {author} {\bibfnamefont {G.}~\bibnamefont {Doolen}},\
  }\bibfield  {title} {\bibinfo {title} {Thermodynamic foundations of kinetic
  theory and lattice {B}oltzmann models for multiphase flows.},\ }\href@noop {}
  {\bibfield  {journal} {\bibinfo  {journal} {J. Stat. Phys.}\ }\textbf
  {\bibinfo {volume} {107}},\ \bibinfo {pages} {309} (\bibinfo {year}
  {2002})}\BibitemShut {NoStop}%
\bibitem [{\citenamefont {Frezzotti}\ \emph {et~al.}(2005)\citenamefont
  {Frezzotti}, \citenamefont {Gibelli},\ and\ \citenamefont
  {Lorenzani}}]{FGL05}%
  \BibitemOpen
  \bibfield  {author} {\bibinfo {author} {\bibfnamefont {A.}~\bibnamefont
  {Frezzotti}}, \bibinfo {author} {\bibfnamefont {L.}~\bibnamefont {Gibelli}},\
  and\ \bibinfo {author} {\bibfnamefont {S.}~\bibnamefont {Lorenzani}},\
  }\bibfield  {title} {\bibinfo {title} {Mean field kinetic theory description
  of evaporation of a fluid into vacuum},\ }\href@noop {} {\bibfield  {journal}
  {\bibinfo  {journal} {Phys. Fluids}\ }\textbf {\bibinfo {volume} {17}},\
  \bibinfo {pages} {012102} (\bibinfo {year} {2005})}\BibitemShut {NoStop}%
\bibitem [{\citenamefont {Benilov}\ and\ \citenamefont {Benilov}(2019)}]{BB19}%
  \BibitemOpen
  \bibfield  {author} {\bibinfo {author} {\bibfnamefont {E.}~\bibnamefont
  {Benilov}}\ and\ \bibinfo {author} {\bibfnamefont {M.}~\bibnamefont
  {Benilov}},\ }\bibfield  {title} {\bibinfo {title} {The {E}nskog--{V}lasov
  equation: a kinetic model describing gas, liquid, and solid},\ }\href@noop {}
  {\bibfield  {journal} {\bibinfo  {journal} {J. Stat. Mech-Theory E}\ }\textbf
  {\bibinfo {volume} {2019}},\ \bibinfo {pages} {103205} (\bibinfo {year}
  {2019})}\BibitemShut {NoStop}%
\bibitem [{\citenamefont {Takata}\ \emph {et~al.}(2018)\citenamefont {Takata},
  \citenamefont {Matsumoto}, \citenamefont {Hirahara},\ and\ \citenamefont
  {Hattori}}]{TMHH18}%
  \BibitemOpen
  \bibfield  {author} {\bibinfo {author} {\bibfnamefont {S.}~\bibnamefont
  {Takata}}, \bibinfo {author} {\bibfnamefont {T.}~\bibnamefont {Matsumoto}},
  \bibinfo {author} {\bibfnamefont {A.}~\bibnamefont {Hirahara}},\ and\
  \bibinfo {author} {\bibfnamefont {M.}~\bibnamefont {Hattori}},\ }\bibfield
  {title} {\bibinfo {title} {Kinetic theory for a simple modeling of a phase
  transition: Dynamics out of local equilibrium},\ }\href@noop {} {\bibfield
  {journal} {\bibinfo  {journal} {Phys. Rev. E}\ }\textbf {\bibinfo {volume}
  {98}},\ \bibinfo {pages} {052123} (\bibinfo {year} {2018})}\BibitemShut
  {NoStop}%
\bibitem [{\citenamefont {Benilov}\ and\ \citenamefont {Benilov}(2018)}]{BB18}%
  \BibitemOpen
  \bibfield  {author} {\bibinfo {author} {\bibfnamefont {E.~S.}\ \bibnamefont
  {Benilov}}\ and\ \bibinfo {author} {\bibfnamefont {M.~S.}\ \bibnamefont
  {Benilov}},\ }\bibfield  {title} {\bibinfo {title} {Energy conservation and
  {H} theorem for the {E}nskog--{V}lasov equation},\ }\href@noop {} {\bibfield
  {journal} {\bibinfo  {journal} {Phys. Rev. E}\ }\textbf {\bibinfo {volume}
  {97}},\ \bibinfo {pages} {062115} (\bibinfo {year} {2018})}\BibitemShut
  {NoStop}%
\bibitem [{\citenamefont {Takata}\ and\ \citenamefont
  {Noguchi}(2018)}]{takata2018waals}%
  \BibitemOpen
  \bibfield  {author} {\bibinfo {author} {\bibfnamefont {S.}~\bibnamefont
  {Takata}}\ and\ \bibinfo {author} {\bibfnamefont {T.}~\bibnamefont
  {Noguchi}},\ }\bibfield  {title} {\bibinfo {title} {Kinetic model for the
  phase transition of the van der {W}aals fluid},\ }\href@noop {} {\bibfield
  {journal} {\bibinfo  {journal} {Journal of Statistical Physics}\ }\textbf
  {\bibinfo {volume} {172}},\ \bibinfo {pages} {880} (\bibinfo {year}
  {2018})}\BibitemShut {NoStop}%
\bibitem [{\citenamefont {Frezzotti}\ \emph {et~al.}(2019)\citenamefont
  {Frezzotti}, \citenamefont {Barbante},\ and\ \citenamefont
  {Gibelli}}]{FBG19}%
  \BibitemOpen
  \bibfield  {author} {\bibinfo {author} {\bibfnamefont {A.}~\bibnamefont
  {Frezzotti}}, \bibinfo {author} {\bibfnamefont {P.}~\bibnamefont
  {Barbante}},\ and\ \bibinfo {author} {\bibfnamefont {L.}~\bibnamefont
  {Gibelli}},\ }\bibfield  {title} {\bibinfo {title} {Direct simulation {M}onte
  {C}arlo applications to liquid-vapor flows},\ }\href@noop {} {\bibfield
  {journal} {\bibinfo  {journal} {Phys. Fluids}\ }\textbf {\bibinfo {volume}
  {31}},\ \bibinfo {pages} {062103} (\bibinfo {year} {2019})}\BibitemShut
  {NoStop}%
\bibitem [{\citenamefont {Busuioc}\ \emph
  {et~al.}(2020{\natexlab{a}})\citenamefont {Busuioc}, \citenamefont {Gibelli},
  \citenamefont {Lockerby},\ and\ \citenamefont {Sprittles}}]{BGLS20}%
  \BibitemOpen
  \bibfield  {author} {\bibinfo {author} {\bibfnamefont {S.}~\bibnamefont
  {Busuioc}}, \bibinfo {author} {\bibfnamefont {L.}~\bibnamefont {Gibelli}},
  \bibinfo {author} {\bibfnamefont {D.~A.}\ \bibnamefont {Lockerby}},\ and\
  \bibinfo {author} {\bibfnamefont {J.~E.}\ \bibnamefont {Sprittles}},\
  }\bibfield  {title} {\bibinfo {title} {Velocity distribution function of
  spontaneously evaporating atoms},\ }\href
  {https://doi.org/10.1103/PhysRevFluids.5.103401} {\bibfield  {journal}
  {\bibinfo  {journal} {Phys. Rev. Fluids}\ }\textbf {\bibinfo {volume} {5}},\
  \bibinfo {pages} {103401} (\bibinfo {year} {2020}{\natexlab{a}})}\BibitemShut
  {NoStop}%
\bibitem [{\citenamefont {Busuioc}\ and\ \citenamefont {Gibelli}(2020)}]{BG20}%
  \BibitemOpen
  \bibfield  {author} {\bibinfo {author} {\bibfnamefont {S.}~\bibnamefont
  {Busuioc}}\ and\ \bibinfo {author} {\bibfnamefont {L.}~\bibnamefont
  {Gibelli}},\ }\bibfield  {title} {\bibinfo {title} {Mean-field kinetic theory
  approach to {L}angmuir evaporation of polyatomic liquids},\ }\href
  {https://doi.org/10.1063/5.0021227} {\bibfield  {journal} {\bibinfo
  {journal} {Physics of Fluids}\ }\textbf {\bibinfo {volume} {32}},\ \bibinfo
  {pages} {093314} (\bibinfo {year} {2020})}\BibitemShut {NoStop}%
\bibitem [{\citenamefont {Takata}\ \emph {et~al.}(2021)\citenamefont {Takata},
  \citenamefont {Matsumoto},\ and\ \citenamefont
  {Hattori}}]{takata2021kinetic}%
  \BibitemOpen
  \bibfield  {author} {\bibinfo {author} {\bibfnamefont {S.}~\bibnamefont
  {Takata}}, \bibinfo {author} {\bibfnamefont {T.}~\bibnamefont {Matsumoto}},\
  and\ \bibinfo {author} {\bibfnamefont {M.}~\bibnamefont {Hattori}},\
  }\bibfield  {title} {\bibinfo {title} {Kinetic model for the phase transition
  of the van der {W}aals fluid},\ }\href@noop {} {\bibfield  {journal}
  {\bibinfo  {journal} {Physical Review E}\ }\textbf {\bibinfo {volume}
  {103}},\ \bibinfo {pages} {062110} (\bibinfo {year} {2021})}\BibitemShut
  {NoStop}%
\bibitem [{\citenamefont {Busuioc}\ \emph {et~al.}(2023)\citenamefont
  {Busuioc}, \citenamefont {Frezzotti},\ and\ \citenamefont {Gibelli}}]{BFG23}%
  \BibitemOpen
  \bibfield  {author} {\bibinfo {author} {\bibfnamefont {S.}~\bibnamefont
  {Busuioc}}, \bibinfo {author} {\bibfnamefont {A.}~\bibnamefont {Frezzotti}},\
  and\ \bibinfo {author} {\bibfnamefont {L.}~\bibnamefont {Gibelli}},\
  }\bibfield  {title} {\bibinfo {title} {A weighted particle scheme for
  {E}nskog-{V}lasov equation to simulate spherical nano-droplets/bubbles},\
  }\href {https://doi.org/https://doi.org/10.1016/j.jcp.2022.111873} {\bibfield
   {journal} {\bibinfo  {journal} {Journal of Computational Physics}\ }\textbf
  {\bibinfo {volume} {475}},\ \bibinfo {pages} {111873} (\bibinfo {year}
  {2023})}\BibitemShut {NoStop}%
\bibitem [{\citenamefont {Chen}\ \emph {et~al.}(2023)\citenamefont {Chen},
  \citenamefont {Wu}, \citenamefont {Wang},\ and\ \citenamefont
  {Chen}}]{CWWC23}%
  \BibitemOpen
  \bibfield  {author} {\bibinfo {author} {\bibfnamefont {T.}~\bibnamefont
  {Chen}}, \bibinfo {author} {\bibfnamefont {L.}~\bibnamefont {Wu}}, \bibinfo
  {author} {\bibfnamefont {L.-P.}\ \bibnamefont {Wang}},\ and\ \bibinfo
  {author} {\bibfnamefont {S.}~\bibnamefont {Chen}},\ }\bibfield  {title}
  {\bibinfo {title} {Rarefaction effects in head-on collision of two
  near-critical droplets},\ }\href
  {https://doi.org/https://doi.org/10.1016/j.ijmultiphaseflow.2023.104451}
  {\bibfield  {journal} {\bibinfo  {journal} {International Journal of
  Multiphase Flow}\ }\textbf {\bibinfo {volume} {164}},\ \bibinfo {pages}
  {104451} (\bibinfo {year} {2023})}\BibitemShut {NoStop}%
\bibitem [{\citenamefont {Wang}\ \emph {et~al.}(2020)\citenamefont {Wang},
  \citenamefont {Wu}, \citenamefont {Ho}, \citenamefont {Li}, \citenamefont
  {Li},\ and\ \citenamefont {Zhang}}]{WWHLLZ20}%
  \BibitemOpen
  \bibfield  {author} {\bibinfo {author} {\bibfnamefont {P.}~\bibnamefont
  {Wang}}, \bibinfo {author} {\bibfnamefont {L.}~\bibnamefont {Wu}}, \bibinfo
  {author} {\bibfnamefont {M.~T.}\ \bibnamefont {Ho}}, \bibinfo {author}
  {\bibfnamefont {J.}~\bibnamefont {Li}}, \bibinfo {author} {\bibfnamefont
  {Z.-H.}\ \bibnamefont {Li}},\ and\ \bibinfo {author} {\bibfnamefont
  {Y.}~\bibnamefont {Zhang}},\ }\bibfield  {title} {\bibinfo {title} {The
  kinetic {S}hakhov–{E}nskog model for non-equilibrium flow of dense gases},\
  }\href {https://doi.org/10.1017/jfm.2019.915} {\bibfield  {journal} {\bibinfo
   {journal} {Journal of Fluid Mechanics}\ }\textbf {\bibinfo {volume} {883}},\
  \bibinfo {pages} {A48} (\bibinfo {year} {2020})}\BibitemShut {NoStop}%
\bibitem [{\citenamefont {Su}\ \emph {et~al.}(2023)\citenamefont {Su},
  \citenamefont {Gibelli}, \citenamefont {Li}, \citenamefont {Borg},\ and\
  \citenamefont {Zhang}}]{WGLBZ23}%
  \BibitemOpen
  \bibfield  {author} {\bibinfo {author} {\bibfnamefont {W.}~\bibnamefont
  {Su}}, \bibinfo {author} {\bibfnamefont {L.}~\bibnamefont {Gibelli}},
  \bibinfo {author} {\bibfnamefont {J.}~\bibnamefont {Li}}, \bibinfo {author}
  {\bibfnamefont {M.~K.}\ \bibnamefont {Borg}},\ and\ \bibinfo {author}
  {\bibfnamefont {Y.}~\bibnamefont {Zhang}},\ }\bibfield  {title} {\bibinfo
  {title} {Kinetic modeling of nonequilibrium flow of hard-sphere dense
  gases},\ }\href {https://doi.org/10.1103/PhysRevFluids.8.013401} {\bibfield
  {journal} {\bibinfo  {journal} {Phys. Rev. Fluids}\ }\textbf {\bibinfo
  {volume} {8}},\ \bibinfo {pages} {013401} (\bibinfo {year}
  {2023})}\BibitemShut {NoStop}%
\bibitem [{\citenamefont {Li}\ \emph {et~al.}(2024)\citenamefont {Li},
  \citenamefont {Su}, \citenamefont {Shan}, \citenamefont {Li}, \citenamefont
  {Gibelli},\ and\ \citenamefont {Zhang}}]{LSSLGZ24}%
  \BibitemOpen
  \bibfield  {author} {\bibinfo {author} {\bibfnamefont {S.}~\bibnamefont
  {Li}}, \bibinfo {author} {\bibfnamefont {W.}~\bibnamefont {Su}}, \bibinfo
  {author} {\bibfnamefont {B.}~\bibnamefont {Shan}}, \bibinfo {author}
  {\bibfnamefont {Z.}~\bibnamefont {Li}}, \bibinfo {author} {\bibfnamefont
  {L.}~\bibnamefont {Gibelli}},\ and\ \bibinfo {author} {\bibfnamefont
  {Y.}~\bibnamefont {Zhang}},\ }\bibfield  {title} {\bibinfo {title} {Molecular
  kinetic modelling of non-equilibrium evaporative flows},\ }\href
  {https://doi.org/10.1017/jfm.2024.605} {\bibfield  {journal} {\bibinfo
  {journal} {Journal of Fluid Mechanics}\ }\textbf {\bibinfo {volume} {994}},\
  \bibinfo {pages} {A16} (\bibinfo {year} {2024})}\BibitemShut {NoStop}%
\bibitem [{\citenamefont {Zhang}\ \emph {et~al.}(2020)\citenamefont {Zhang},
  \citenamefont {Xu}, \citenamefont {Qiu}, \citenamefont {Wei},\ and\
  \citenamefont {Wei}}]{ZXQWW20}%
  \BibitemOpen
  \bibfield  {author} {\bibinfo {author} {\bibfnamefont {Y.-D.}\ \bibnamefont
  {Zhang}}, \bibinfo {author} {\bibfnamefont {A.-G.}\ \bibnamefont {Xu}},
  \bibinfo {author} {\bibfnamefont {J.-J.}\ \bibnamefont {Qiu}}, \bibinfo
  {author} {\bibfnamefont {H.-T.}\ \bibnamefont {Wei}},\ and\ \bibinfo {author}
  {\bibfnamefont {Z.-H.}\ \bibnamefont {Wei}},\ }\bibfield  {title} {\bibinfo
  {title} {Kinetic modeling of multiphase flow based on simplified {E}nskog
  equation.},\ }\href@noop {} {\bibfield  {journal} {\bibinfo  {journal}
  {Front. Phys.}\ }\textbf {\bibinfo {volume} {15}},\ \bibinfo {pages} {62503}
  (\bibinfo {year} {2020})}\BibitemShut {NoStop}%
\bibitem [{\citenamefont {Gan}\ \emph {et~al.}(2022)\citenamefont {Gan},
  \citenamefont {Xu}, \citenamefont {Lai}, \citenamefont {Li}, \citenamefont
  {Sun},\ and\ \citenamefont {Succi}}]{GXLLSS22}%
  \BibitemOpen
  \bibfield  {author} {\bibinfo {author} {\bibfnamefont {Y.}~\bibnamefont
  {Gan}}, \bibinfo {author} {\bibfnamefont {A.}~\bibnamefont {Xu}}, \bibinfo
  {author} {\bibfnamefont {H.}~\bibnamefont {Lai}}, \bibinfo {author}
  {\bibfnamefont {W.}~\bibnamefont {Li}}, \bibinfo {author} {\bibfnamefont
  {G.}~\bibnamefont {Sun}},\ and\ \bibinfo {author} {\bibfnamefont
  {S.}~\bibnamefont {Succi}},\ }\bibfield  {title} {\bibinfo {title} {Discrete
  {B}oltzmann multi-scale modelling of non-equilibrium multiphase flows},\
  }\href {https://doi.org/10.1017/jfm.2022.844} {\bibfield  {journal} {\bibinfo
   {journal} {Journal of Fluid Mechanics}\ }\textbf {\bibinfo {volume} {951}},\
  \bibinfo {pages} {A8} (\bibinfo {year} {2022})}\BibitemShut {NoStop}%
\bibitem [{\citenamefont {Huang}\ \emph {et~al.}(2021)\citenamefont {Huang},
  \citenamefont {Wu},\ and\ \citenamefont {Adams}}]{HWA21}%
  \BibitemOpen
  \bibfield  {author} {\bibinfo {author} {\bibfnamefont {R.}~\bibnamefont
  {Huang}}, \bibinfo {author} {\bibfnamefont {H.}~\bibnamefont {Wu}},\ and\
  \bibinfo {author} {\bibfnamefont {N.~A.}\ \bibnamefont {Adams}},\ }\bibfield
  {title} {\bibinfo {title} {Mesoscopic lattice {B}oltzmann modeling of the
  liquid-vapor phase transition},\ }\href
  {https://doi.org/10.1103/PhysRevLett.126.244501} {\bibfield  {journal}
  {\bibinfo  {journal} {Phys. Rev. Lett.}\ }\textbf {\bibinfo {volume} {126}},\
  \bibinfo {pages} {244501} (\bibinfo {year} {2021})}\BibitemShut {NoStop}%
\bibitem [{\citenamefont {Busuioc}(2023)}]{B23}%
  \BibitemOpen
  \bibfield  {author} {\bibinfo {author} {\bibfnamefont {S.}~\bibnamefont
  {Busuioc}},\ }\bibfield  {title} {\bibinfo {title} {Quadrature-based lattice
  {B}oltzmann model for non-equilibrium dense gas flows},\ }\href
  {https://doi.org/10.1063/5.0135579} {\bibfield  {journal} {\bibinfo
  {journal} {Physics of Fluids}\ }\textbf {\bibinfo {volume} {35}},\ \bibinfo
  {pages} {016112} (\bibinfo {year} {2023})}\BibitemShut {NoStop}%
\bibitem [{\citenamefont {Busuioc}\ and\ \citenamefont {Sofonea}(2024)}]{BS24}%
  \BibitemOpen
  \bibfield  {author} {\bibinfo {author} {\bibfnamefont {S.}~\bibnamefont
  {Busuioc}}\ and\ \bibinfo {author} {\bibfnamefont {V.}~\bibnamefont
  {Sofonea}},\ }\bibfield  {title} {\bibinfo {title} {Bounded flows of dense
  gases},\ }\href {https://doi.org/10.1103/PhysRevFluids.9.023401} {\bibfield
  {journal} {\bibinfo  {journal} {Phys. Rev. Fluids}\ }\textbf {\bibinfo
  {volume} {9}},\ \bibinfo {pages} {023401} (\bibinfo {year}
  {2024})}\BibitemShut {NoStop}%
\bibitem [{\citenamefont {Busuioc}(2024)}]{B24}%
  \BibitemOpen
  \bibfield  {author} {\bibinfo {author} {\bibfnamefont {S.}~\bibnamefont
  {Busuioc}},\ }\bibfield  {title} {\bibinfo {title} {Mesoscopic lattice
  {B}oltzmann modeling of dense gas flows in curvilinear geometries},\ }\href
  {https://doi.org/10.1103/PhysRevFluids.9.053401} {\bibfield  {journal}
  {\bibinfo  {journal} {Phys. Rev. Fluids}\ }\textbf {\bibinfo {volume} {9}},\
  \bibinfo {pages} {053401} (\bibinfo {year} {2024})}\BibitemShut {NoStop}%
\bibitem [{\citenamefont {Shan}\ \emph {et~al.}(2006)\citenamefont {Shan},
  \citenamefont {Yuan},\ and\ \citenamefont {Chen}}]{SYC06}%
  \BibitemOpen
  \bibfield  {author} {\bibinfo {author} {\bibfnamefont {X.}~\bibnamefont
  {Shan}}, \bibinfo {author} {\bibfnamefont {X.-F.}\ \bibnamefont {Yuan}},\
  and\ \bibinfo {author} {\bibfnamefont {H.}~\bibnamefont {Chen}},\ }\bibfield
  {title} {\bibinfo {title} {Kinetic theory representation of hydrodynamics: a
  way beyond the {N}avier–{S}tokes equation},\ }\href
  {https://doi.org/10.1017/S0022112005008153} {\bibfield  {journal} {\bibinfo
  {journal} {Journal of Fluid Mechanics}\ }\textbf {\bibinfo {volume} {550}},\
  \bibinfo {pages} {413} (\bibinfo {year} {2006})}\BibitemShut {NoStop}%
\bibitem [{\citenamefont {Ambruş}\ and\ \citenamefont
  {Sofonea}(2016{\natexlab{a}})}]{AS16a}%
  \BibitemOpen
  \bibfield  {author} {\bibinfo {author} {\bibfnamefont {V.}~\bibnamefont
  {Ambruş}}\ and\ \bibinfo {author} {\bibfnamefont {V.}~\bibnamefont
  {Sofonea}},\ }\bibfield  {title} {\bibinfo {title} {{Lattice {B}oltzmann
  models based on half-range Gauss-Hermite quadratures}},\ }\href
  {https://doi.org/http://dx.doi.org/10.1016/j.jcp.2016.04.010} {\bibfield
  {journal} {\bibinfo  {journal} {J. Comput. Phys.}\ }\textbf {\bibinfo
  {volume} {316}},\ \bibinfo {pages} {760} (\bibinfo {year}
  {2016}{\natexlab{a}})}\BibitemShut {NoStop}%
\bibitem [{\citenamefont {Ambruş}\ and\ \citenamefont
  {Sofonea}(2016{\natexlab{b}})}]{AS16b}%
  \BibitemOpen
  \bibfield  {author} {\bibinfo {author} {\bibfnamefont {V.}~\bibnamefont
  {Ambruş}}\ and\ \bibinfo {author} {\bibfnamefont {V.}~\bibnamefont
  {Sofonea}},\ }\bibfield  {title} {\bibinfo {title} {{Application of mixed
  quadrature lattice {B}oltzmann models for the simulation of Poiseuille flow
  at non-negligible values of the Knudsen number}},\ }\href
  {https://doi.org/http://dx.doi.org/10.1016/j.jocs.2016.03.016} {\bibfield
  {journal} {\bibinfo  {journal} {J. Comput. Science}\ }\textbf {\bibinfo
  {volume} {17}},\ \bibinfo {pages} {403} (\bibinfo {year}
  {2016}{\natexlab{b}})}\BibitemShut {NoStop}%
\bibitem [{\citenamefont {Ambruș}\ and\ \citenamefont
  {Sofonea}(2019)}]{AS2019}%
  \BibitemOpen
  \bibfield  {author} {\bibinfo {author} {\bibfnamefont {V.~E.}\ \bibnamefont
  {Ambruș}}\ and\ \bibinfo {author} {\bibfnamefont {V.}~\bibnamefont
  {Sofonea}},\ }\bibinfo {title} {Quadrature-based lattice {B}oltzmann models
  for rarefied gas flow},\ in\ \href@noop {} {\emph {\bibinfo {booktitle}
  {Flowing Matter}}},\ \bibinfo {editor} {edited by\ \bibinfo {editor}
  {\bibfnamefont {F.}~\bibnamefont {Toschi}}\ and\ \bibinfo {editor}
  {\bibfnamefont {M.}~\bibnamefont {Sega}}}\ (\bibinfo  {publisher} {Springer
  International Publishing},\ \bibinfo {address} {Cham},\ \bibinfo {year}
  {2019})\ pp.\ \bibinfo {pages} {271--299}\BibitemShut {NoStop}%
\bibitem [{\citenamefont {Busuioc}\ and\ \citenamefont
  {Ambru\ifmmode~\mbox{\c{s}}\else \c{s}\fi{}}(2019)}]{BA19}%
  \BibitemOpen
  \bibfield  {author} {\bibinfo {author} {\bibfnamefont {S.}~\bibnamefont
  {Busuioc}}\ and\ \bibinfo {author} {\bibfnamefont {V.~E.}\ \bibnamefont
  {Ambru\ifmmode~\mbox{\c{s}}\else \c{s}\fi{}}},\ }\bibfield  {title} {\bibinfo
  {title} {Lattice {B}oltzmann models based on the vielbein formalism for the
  simulation of flows in curvilinear geometries},\ }\href
  {https://doi.org/10.1103/PhysRevE.99.033304} {\bibfield  {journal} {\bibinfo
  {journal} {Phys. Rev. E}\ }\textbf {\bibinfo {volume} {99}},\ \bibinfo
  {pages} {033304} (\bibinfo {year} {2019})}\BibitemShut {NoStop}%
\bibitem [{\citenamefont {Carnahan}\ and\ \citenamefont
  {Starling}(1969)}]{CS69}%
  \BibitemOpen
  \bibfield  {author} {\bibinfo {author} {\bibfnamefont {N.~F.}\ \bibnamefont
  {Carnahan}}\ and\ \bibinfo {author} {\bibfnamefont {K.~E.}\ \bibnamefont
  {Starling}},\ }\bibfield  {title} {\bibinfo {title} {Equation of state for
  nonattracting rigid spheres},\ }\href@noop {} {\bibfield  {journal} {\bibinfo
   {journal} {J. Chem. Phys.}\ }\textbf {\bibinfo {volume} {51}},\ \bibinfo
  {pages} {635} (\bibinfo {year} {1969})}\BibitemShut {NoStop}%
\bibitem [{\citenamefont {Fischer}\ and\ \citenamefont
  {Methfessel}(1980)}]{FM80}%
  \BibitemOpen
  \bibfield  {author} {\bibinfo {author} {\bibfnamefont {J.}~\bibnamefont
  {Fischer}}\ and\ \bibinfo {author} {\bibfnamefont {M.}~\bibnamefont
  {Methfessel}},\ }\bibfield  {title} {\bibinfo {title} {{B}orn-{G}reen-{Y}von
  approach to the local densities of a fluid at interfaces},\ }\href@noop {}
  {\bibfield  {journal} {\bibinfo  {journal} {Phys. Rev. A}\ }\textbf {\bibinfo
  {volume} {22}},\ \bibinfo {pages} {2836} (\bibinfo {year}
  {1980})}\BibitemShut {NoStop}%
\bibitem [{\citenamefont {Kon}\ \emph {et~al.}(2014)\citenamefont {Kon},
  \citenamefont {Kobayashi},\ and\ \citenamefont {Watanabe}}]{KKW14}%
  \BibitemOpen
  \bibfield  {author} {\bibinfo {author} {\bibfnamefont {M.}~\bibnamefont
  {Kon}}, \bibinfo {author} {\bibfnamefont {K.}~\bibnamefont {Kobayashi}},\
  and\ \bibinfo {author} {\bibfnamefont {M.}~\bibnamefont {Watanabe}},\
  }\bibfield  {title} {\bibinfo {title} {Method of determining kinetic boundary
  conditions in net evaporation/condensation},\ }\href@noop {} {\bibfield
  {journal} {\bibinfo  {journal} {Phys. Fluids}\ }\textbf {\bibinfo {volume}
  {26}},\ \bibinfo {pages} {072003} (\bibinfo {year} {2014})}\BibitemShut
  {NoStop}%
\bibitem [{\citenamefont {Bruno}\ and\ \citenamefont
  {Frezzotti}(2019)}]{Bruno2019}%
  \BibitemOpen
  \bibfield  {author} {\bibinfo {author} {\bibfnamefont {D.}~\bibnamefont
  {Bruno}}\ and\ \bibinfo {author} {\bibfnamefont {A.}~\bibnamefont
  {Frezzotti}},\ }\bibfield  {title} {\bibinfo {title} {{Dense gas effects in
  the Rayleigh-Brillouin scattering spectra of SF6}},\ }\href@noop {}
  {\bibfield  {journal} {\bibinfo  {journal} {Chem. Phys. Lett.}\ }\textbf
  {\bibinfo {volume} {731}},\ \bibinfo {pages} {136595} (\bibinfo {year}
  {2019})}\BibitemShut {NoStop}%
\bibitem [{\citenamefont {Kobayashi}\ \emph {et~al.}(2017)\citenamefont
  {Kobayashi}, \citenamefont {Sasaki}, \citenamefont {Kon}, \citenamefont
  {Fujii},\ and\ \citenamefont {Watanabe}}]{KSKFW17}%
  \BibitemOpen
  \bibfield  {author} {\bibinfo {author} {\bibfnamefont {K.}~\bibnamefont
  {Kobayashi}}, \bibinfo {author} {\bibfnamefont {K.}~\bibnamefont {Sasaki}},
  \bibinfo {author} {\bibfnamefont {M.}~\bibnamefont {Kon}}, \bibinfo {author}
  {\bibfnamefont {H.}~\bibnamefont {Fujii}},\ and\ \bibinfo {author}
  {\bibfnamefont {M.}~\bibnamefont {Watanabe}},\ }\bibfield  {title} {\bibinfo
  {title} {Kinetic boundary conditions for vapor--gas binary mixture},\
  }\href@noop {} {\bibfield  {journal} {\bibinfo  {journal} {Microfluid.
  Nanofluid.}\ }\textbf {\bibinfo {volume} {21}},\ \bibinfo {pages} {53}
  (\bibinfo {year} {2017})}\BibitemShut {NoStop}%
\bibitem [{\citenamefont {Barbante}\ \emph {et~al.}(2015)\citenamefont
  {Barbante}, \citenamefont {Frezzotti},\ and\ \citenamefont
  {Gibelli}}]{BFG15}%
  \BibitemOpen
  \bibfield  {author} {\bibinfo {author} {\bibfnamefont {P.}~\bibnamefont
  {Barbante}}, \bibinfo {author} {\bibfnamefont {A.}~\bibnamefont
  {Frezzotti}},\ and\ \bibinfo {author} {\bibfnamefont {L.}~\bibnamefont
  {Gibelli}},\ }\bibfield  {title} {\bibinfo {title} {A kinetic theory
  description of liquid menisci at the microscale},\ }\href@noop {} {\bibfield
  {journal} {\bibinfo  {journal} {Kinet. Relat. Mod.}\ }\textbf {\bibinfo
  {volume} {8}},\ \bibinfo {pages} {235} (\bibinfo {year} {2015})}\BibitemShut
  {NoStop}%
\bibitem [{\citenamefont {Kremer}(2010)}]{K10}%
  \BibitemOpen
  \bibfield  {author} {\bibinfo {author} {\bibfnamefont {G.~M.}\ \bibnamefont
  {Kremer}},\ }\href@noop {} {\emph {\bibinfo {title} {An introduction to the
  {B}oltzmann equation and transport processes in gases}}}\ (\bibinfo
  {publisher} {Springer-Verlag, Berlin Heidelberg},\ \bibinfo {year}
  {2010})\BibitemShut {NoStop}%
\bibitem [{\citenamefont {Shakhov}(1968{\natexlab{a}})}]{shakhov68a}%
  \BibitemOpen
  \bibfield  {author} {\bibinfo {author} {\bibfnamefont {E.}~\bibnamefont
  {Shakhov}},\ }\bibfield  {title} {\bibinfo {title} {Generalization of the
  {K}rook kinetic relaxation equation},\ }\href
  {https://doi.org/10.1007/BF01029546} {\bibfield  {journal} {\bibinfo
  {journal} {Fluid Dynamics}\ }\textbf {\bibinfo {volume} {3}},\ \bibinfo
  {pages} {95 } (\bibinfo {year} {1968}{\natexlab{a}})}\BibitemShut {NoStop}%
\bibitem [{\citenamefont {Shakhov}(1968{\natexlab{b}})}]{shakhov68b}%
  \BibitemOpen
  \bibfield  {author} {\bibinfo {author} {\bibfnamefont {E.}~\bibnamefont
  {Shakhov}},\ }\bibfield  {title} {\bibinfo {title} {Approximate kinetic
  equations in rarefied gas theory},\ }\href
  {https://doi.org/10.1007/BF01016254} {\bibfield  {journal} {\bibinfo
  {journal} {Fluid Dynamics}\ }\textbf {\bibinfo {volume} {3}},\ \bibinfo
  {pages} {112 – 115} (\bibinfo {year} {1968}{\natexlab{b}})}\BibitemShut
  {NoStop}%
\bibitem [{\citenamefont {Graur}\ and\ \citenamefont
  {Polikarpov}(2009)}]{GP09}%
  \BibitemOpen
  \bibfield  {author} {\bibinfo {author} {\bibfnamefont {I.}~\bibnamefont
  {Graur}}\ and\ \bibinfo {author} {\bibfnamefont {A.}~\bibnamefont
  {Polikarpov}},\ }\bibfield  {title} {\bibinfo {title} {Comparison of
  different kinetic models for the heat transfer problem},\ }\href@noop {}
  {\bibfield  {journal} {\bibinfo  {journal} {Heat Mass Transfer}\ }\textbf
  {\bibinfo {volume} {46}},\ \bibinfo {pages} {237} (\bibinfo {year}
  {2009})}\BibitemShut {NoStop}%
\bibitem [{\citenamefont {Ambru\c{s}}\ \emph {et~al.}(2020)\citenamefont
  {Ambru\c{s}}, \citenamefont {Sharipov},\ and\ \citenamefont
  {Sofonea}}]{ASS20}%
  \BibitemOpen
  \bibfield  {author} {\bibinfo {author} {\bibfnamefont {V.~E.}\ \bibnamefont
  {Ambru\c{s}}}, \bibinfo {author} {\bibfnamefont {F.}~\bibnamefont
  {Sharipov}},\ and\ \bibinfo {author} {\bibfnamefont {V.}~\bibnamefont
  {Sofonea}},\ }\bibfield  {title} {\bibinfo {title} {Comparison of the
  {S}hakhov and ellipsoidal models for the {B}oltzmann equation and {DSMC} for
  ab initio-based particle interactions},\ }\href
  {https://doi.org/https://doi.org/10.1016/j.compfluid.2020.104637} {\bibfield
  {journal} {\bibinfo  {journal} {Computers \& Fluids}\ }\textbf {\bibinfo
  {volume} {211}},\ \bibinfo {pages} {104637} (\bibinfo {year}
  {2020})}\BibitemShut {NoStop}%
\bibitem [{\citenamefont {Sharipov}(2002)}]{S02}%
  \BibitemOpen
  \bibfield  {author} {\bibinfo {author} {\bibfnamefont {F.}~\bibnamefont
  {Sharipov}},\ }\bibfield  {title} {\bibinfo {title} {Application of the
  {C}ercignani–{L}ampis scattering kernel to calculations of rarefied gas
  flows. i. {P}lane flow between two parallel plates},\ }\href
  {https://doi.org/https://doi.org/10.1016/S0997-7546(01)01160-8} {\bibfield
  {journal} {\bibinfo  {journal} {European Journal of Mechanics - B/Fluids}\
  }\textbf {\bibinfo {volume} {21}},\ \bibinfo {pages} {113} (\bibinfo {year}
  {2002})}\BibitemShut {NoStop}%
\bibitem [{\citenamefont {Sharipov}(2003)}]{S03}%
  \BibitemOpen
  \bibfield  {author} {\bibinfo {author} {\bibfnamefont {F.}~\bibnamefont
  {Sharipov}},\ }\bibfield  {title} {\bibinfo {title} {Application of the
  {C}ercignani–{L}ampis scattering kernel to calculations of rarefied gas
  flows. ii. {S}lip and jump coefficients},\ }\href
  {https://doi.org/https://doi.org/10.1016/S0997-7546(03)00017-7} {\bibfield
  {journal} {\bibinfo  {journal} {European Journal of Mechanics - B/Fluids}\
  }\textbf {\bibinfo {volume} {22}},\ \bibinfo {pages} {133} (\bibinfo {year}
  {2003})}\BibitemShut {NoStop}%
\bibitem [{\citenamefont {Ambru\ifmmode~\mbox{\c{s}}\else \c{s}\fi{}}\ and\
  \citenamefont {Sofonea}(2018)}]{AS18}%
  \BibitemOpen
  \bibfield  {author} {\bibinfo {author} {\bibfnamefont {V.~E.}\ \bibnamefont
  {Ambru\ifmmode~\mbox{\c{s}}\else \c{s}\fi{}}}\ and\ \bibinfo {author}
  {\bibfnamefont {V.}~\bibnamefont {Sofonea}},\ }\bibfield  {title} {\bibinfo
  {title} {{Half-range lattice {B}oltzmann models for the simulation of Couette
  flow using the Shakhov collision term}},\ }\href
  {https://doi.org/10.1103/PhysRevE.98.063311} {\bibfield  {journal} {\bibinfo
  {journal} {Phys. Rev. E}\ }\textbf {\bibinfo {volume} {98}},\ \bibinfo
  {pages} {063311} (\bibinfo {year} {2018})}\BibitemShut {NoStop}%
\bibitem [{\citenamefont {Zhang}\ \emph {et~al.}(2019)\citenamefont {Zhang},
  \citenamefont {Xu}, \citenamefont {Zhang}, \citenamefont {Chen},\ and\
  \citenamefont {Wang}}]{ZXZCW19}%
  \BibitemOpen
  \bibfield  {author} {\bibinfo {author} {\bibfnamefont {Y.}~\bibnamefont
  {Zhang}}, \bibinfo {author} {\bibfnamefont {A.}~\bibnamefont {Xu}}, \bibinfo
  {author} {\bibfnamefont {G.}~\bibnamefont {Zhang}}, \bibinfo {author}
  {\bibfnamefont {Z.}~\bibnamefont {Chen}},\ and\ \bibinfo {author}
  {\bibfnamefont {P.}~\bibnamefont {Wang}},\ }\bibfield  {title} {\bibinfo
  {title} {Discrete {B}oltzmann method for non-equilibrium flows: Based on
  {S}hakhov model},\ }\href
  {https://doi.org/https://doi.org/10.1016/j.cpc.2018.12.018} {\bibfield
  {journal} {\bibinfo  {journal} {Computer Physics Communications}\ }\textbf
  {\bibinfo {volume} {238}},\ \bibinfo {pages} {50} (\bibinfo {year}
  {2019})}\BibitemShut {NoStop}%
\bibitem [{\citenamefont {Todorova}\ \emph {et~al.}(2020)\citenamefont
  {Todorova}, \citenamefont {White},\ and\ \citenamefont {Steijl}}]{TWS20}%
  \BibitemOpen
  \bibfield  {author} {\bibinfo {author} {\bibfnamefont {B.~N.}\ \bibnamefont
  {Todorova}}, \bibinfo {author} {\bibfnamefont {C.}~\bibnamefont {White}},\
  and\ \bibinfo {author} {\bibfnamefont {R.}~\bibnamefont {Steijl}},\
  }\bibfield  {title} {\bibinfo {title} {Modeling of nitrogen and oxygen gas
  mixture with a novel diatomic kinetic model},\ }\href
  {https://doi.org/10.1063/5.0021672} {\bibfield  {journal} {\bibinfo
  {journal} {AIP Advances}\ }\textbf {\bibinfo {volume} {10}},\ \bibinfo
  {pages} {095218} (\bibinfo {year} {2020})}\BibitemShut {NoStop}%
\bibitem [{\citenamefont {Struchtrup}\ and\ \citenamefont
  {Frezzotti}(2022)}]{SF22}%
  \BibitemOpen
  \bibfield  {author} {\bibinfo {author} {\bibfnamefont {H.}~\bibnamefont
  {Struchtrup}}\ and\ \bibinfo {author} {\bibfnamefont {A.}~\bibnamefont
  {Frezzotti}},\ }\bibfield  {title} {\bibinfo {title} {Twenty-six moment
  equations for the {E}nskog–{V}lasov equation},\ }\href
  {https://doi.org/10.1017/jfm.2022.98} {\bibfield  {journal} {\bibinfo
  {journal} {Journal of Fluid Mechanics}\ }\textbf {\bibinfo {volume} {940}},\
  \bibinfo {pages} {A40} (\bibinfo {year} {2022})}\BibitemShut {NoStop}%
\bibitem [{\citenamefont {Balescu}(1975)}]{B75}%
  \BibitemOpen
  \bibfield  {author} {\bibinfo {author} {\bibfnamefont {R.}~\bibnamefont
  {Balescu}},\ }\href@noop {} {\emph {\bibinfo {title} {Equilibrium and
  nonequilibrium statistical mechanics}}}\ (\bibinfo  {publisher} {John Wiley,
  New York},\ \bibinfo {year} {1975})\BibitemShut {NoStop}%
\bibitem [{\citenamefont {Negro}\ \emph {et~al.}(2019)\citenamefont {Negro},
  \citenamefont {Busuioc}, \citenamefont {Ambruș}, \citenamefont {Gonnella},
  \citenamefont {Lamura},\ and\ \citenamefont {Sofonea}}]{NBAGLS19}%
  \BibitemOpen
  \bibfield  {author} {\bibinfo {author} {\bibfnamefont {G.}~\bibnamefont
  {Negro}}, \bibinfo {author} {\bibfnamefont {S.}~\bibnamefont {Busuioc}},
  \bibinfo {author} {\bibfnamefont {V.~E.}\ \bibnamefont {Ambruș}}, \bibinfo
  {author} {\bibfnamefont {G.}~\bibnamefont {Gonnella}}, \bibinfo {author}
  {\bibfnamefont {A.}~\bibnamefont {Lamura}},\ and\ \bibinfo {author}
  {\bibfnamefont {V.}~\bibnamefont {Sofonea}},\ }\bibfield  {title} {\bibinfo
  {title} {Comparison between isothermal collision-streaming and
  finite-difference lattice {B}oltzmann models},\ }\bibfield  {journal}
  {\bibinfo  {journal} {{I}nternational {J}ournal of {M}odern {P}hysics {C}}\
  }\textbf {\bibinfo {volume} {30}},\ \href
  {https://doi.org/10.1142/S0129183119410055} {10.1142/S0129183119410055}
  (\bibinfo {year} {2019})\BibitemShut {NoStop}%
\bibitem [{\citenamefont {Hirschfelder}\ \emph {et~al.}(1964)\citenamefont
  {Hirschfelder}, \citenamefont {Bird},\ and\ \citenamefont {Curtiss}}]{HBC64}%
  \BibitemOpen
  \bibfield  {author} {\bibinfo {author} {\bibfnamefont {J.}~\bibnamefont
  {Hirschfelder}}, \bibinfo {author} {\bibfnamefont {R.~B.}\ \bibnamefont
  {Bird}},\ and\ \bibinfo {author} {\bibfnamefont {C.~F.}\ \bibnamefont
  {Curtiss}},\ }\href@noop {} {\emph {\bibinfo {title} {Molecular theory of
  gases and liquids}}}\ (\bibinfo  {publisher} {Wiley},\ \bibinfo {year}
  {1964})\BibitemShut {NoStop}%
\bibitem [{\citenamefont {Ambru\ifmmode~\mbox{\c{s}}\else \c{s}\fi{}}\ \emph
  {et~al.}(2019)\citenamefont {Ambru\ifmmode~\mbox{\c{s}}\else \c{s}\fi{}},
  \citenamefont {Busuioc}, \citenamefont {Wagner}, \citenamefont {Paillusson},\
  and\ \citenamefont {Kusumaatmaja}}]{ABWPK19}%
  \BibitemOpen
  \bibfield  {author} {\bibinfo {author} {\bibfnamefont {V.~E.}\ \bibnamefont
  {Ambru\ifmmode~\mbox{\c{s}}\else \c{s}\fi{}}}, \bibinfo {author}
  {\bibfnamefont {S.}~\bibnamefont {Busuioc}}, \bibinfo {author} {\bibfnamefont
  {A.~J.}\ \bibnamefont {Wagner}}, \bibinfo {author} {\bibfnamefont
  {F.}~\bibnamefont {Paillusson}},\ and\ \bibinfo {author} {\bibfnamefont
  {H.}~\bibnamefont {Kusumaatmaja}},\ }\bibfield  {title} {\bibinfo {title}
  {Multicomponent flow on curved surfaces: A vielbein lattice {B}oltzmann
  approach},\ }\href {https://doi.org/10.1103/PhysRevE.100.063306} {\bibfield
  {journal} {\bibinfo  {journal} {Phys. Rev. E}\ }\textbf {\bibinfo {volume}
  {100}},\ \bibinfo {pages} {063306} (\bibinfo {year} {2019})}\BibitemShut
  {NoStop}%
\bibitem [{\citenamefont {Sofonea}\ \emph {et~al.}(2018)\citenamefont
  {Sofonea}, \citenamefont {Biciu\ifmmode \mbox{\c{s}}\else
  \c{s}\fi{}c\ifmmode~\u{a}\else \u{a}\fi{}}, \citenamefont {Busuioc},
  \citenamefont {Ambru\ifmmode~\mbox{\c{s}}\else \c{s}\fi{}}, \citenamefont
  {Gonnella},\ and\ \citenamefont {Lamura}}]{SBBAGL18}%
  \BibitemOpen
  \bibfield  {author} {\bibinfo {author} {\bibfnamefont {V.}~\bibnamefont
  {Sofonea}}, \bibinfo {author} {\bibfnamefont {T.}~\bibnamefont {Biciu\ifmmode
  \mbox{\c{s}}\else \c{s}\fi{}c\ifmmode~\u{a}\else \u{a}\fi{}}}, \bibinfo
  {author} {\bibfnamefont {S.}~\bibnamefont {Busuioc}}, \bibinfo {author}
  {\bibfnamefont {V.~E.}\ \bibnamefont {Ambru\ifmmode~\mbox{\c{s}}\else
  \c{s}\fi{}}}, \bibinfo {author} {\bibfnamefont {G.}~\bibnamefont
  {Gonnella}},\ and\ \bibinfo {author} {\bibfnamefont {A.}~\bibnamefont
  {Lamura}},\ }\bibfield  {title} {\bibinfo {title} {Corner-transport-upwind
  lattice {B}oltzmann model for bubble cavitation},\ }\href
  {https://doi.org/10.1103/PhysRevE.97.023309} {\bibfield  {journal} {\bibinfo
  {journal} {Phys. Rev. E}\ }\textbf {\bibinfo {volume} {97}},\ \bibinfo
  {pages} {023309} (\bibinfo {year} {2018})}\BibitemShut {NoStop}%
\bibitem [{\citenamefont {Busuioc}\ \emph
  {et~al.}(2020{\natexlab{b}})\citenamefont {Busuioc}, \citenamefont {Ambruş},
  \citenamefont {Biciuşcă},\ and\ \citenamefont {Sofonea}}]{BABS20}%
  \BibitemOpen
  \bibfield  {author} {\bibinfo {author} {\bibfnamefont {S.}~\bibnamefont
  {Busuioc}}, \bibinfo {author} {\bibfnamefont {V.~E.}\ \bibnamefont
  {Ambruş}}, \bibinfo {author} {\bibfnamefont {T.}~\bibnamefont
  {Biciuşcă}},\ and\ \bibinfo {author} {\bibfnamefont {V.}~\bibnamefont
  {Sofonea}},\ }\bibfield  {title} {\bibinfo {title} {Two-dimensional
  off-lattice {B}oltzmann model for van der {W}aals fluids with variable
  temperature},\ }\href
  {https://doi.org/https://doi.org/10.1016/j.camwa.2018.12.015} {\bibfield
  {journal} {\bibinfo  {journal} {Computers \& Mathematics with Applications}\
  }\textbf {\bibinfo {volume} {79}},\ \bibinfo {pages} {111} (\bibinfo {year}
  {2020}{\natexlab{b}})}\BibitemShut {NoStop}%
\bibitem [{\citenamefont {Shu}\ and\ \citenamefont {Osher}(1988)}]{SO88}%
  \BibitemOpen
  \bibfield  {author} {\bibinfo {author} {\bibfnamefont {C.-W.}\ \bibnamefont
  {Shu}}\ and\ \bibinfo {author} {\bibfnamefont {S.}~\bibnamefont {Osher}},\
  }\bibfield  {title} {\bibinfo {title} {Efficient implementation of
  essentially non-oscillatory shock-capturing schemes},\ }\href
  {https://doi.org/https://doi.org/10.1016/0021-9991(88)90177-5} {\bibfield
  {journal} {\bibinfo  {journal} {Journal of Computational Physics}\ }\textbf
  {\bibinfo {volume} {77}},\ \bibinfo {pages} {439} (\bibinfo {year}
  {1988})}\BibitemShut {NoStop}%
\bibitem [{\citenamefont {Gan}\ \emph {et~al.}(2011)\citenamefont {Gan},
  \citenamefont {Xu}, \citenamefont {Zhang},\ and\ \citenamefont
  {Li}}]{GXZL11}%
  \BibitemOpen
  \bibfield  {author} {\bibinfo {author} {\bibfnamefont {Y.}~\bibnamefont
  {Gan}}, \bibinfo {author} {\bibfnamefont {A.}~\bibnamefont {Xu}}, \bibinfo
  {author} {\bibfnamefont {G.}~\bibnamefont {Zhang}},\ and\ \bibinfo {author}
  {\bibfnamefont {Y.}~\bibnamefont {Li}},\ }\bibfield  {title} {\bibinfo
  {title} {Lattice {B}oltzmann study on {K}elvin-{H}elmholtz instability:
  {R}oles of velocity and density gradients},\ }\href
  {https://doi.org/10.1103/PhysRevE.83.056704} {\bibfield  {journal} {\bibinfo
  {journal} {Phys. Rev. E}\ }\textbf {\bibinfo {volume} {83}},\ \bibinfo
  {pages} {056704} (\bibinfo {year} {2011})}\BibitemShut {NoStop}%
\bibitem [{\citenamefont {Jiang}\ and\ \citenamefont {Shu}(1996)}]{JS96}%
  \BibitemOpen
  \bibfield  {author} {\bibinfo {author} {\bibfnamefont {G.-S.}\ \bibnamefont
  {Jiang}}\ and\ \bibinfo {author} {\bibfnamefont {C.-W.}\ \bibnamefont
  {Shu}},\ }\bibfield  {title} {\bibinfo {title} {Efficient implementation of
  {W}eighted {ENO} schemes},\ }\href
  {https://doi.org/https://doi.org/10.1006/jcph.1996.0130} {\bibfield
  {journal} {\bibinfo  {journal} {Journal of Computational Physics}\ }\textbf
  {\bibinfo {volume} {126}},\ \bibinfo {pages} {202} (\bibinfo {year}
  {1996})}\BibitemShut {NoStop}%
\bibitem [{\citenamefont {Fornberg}(1988)}]{F88}%
  \BibitemOpen
  \bibfield  {author} {\bibinfo {author} {\bibfnamefont {B.}~\bibnamefont
  {Fornberg}},\ }\bibfield  {title} {\bibinfo {title} {Generation of finite
  difference formulas on arbitrarily spaced grids},\ }\href@noop {} {\bibfield
  {journal} {\bibinfo  {journal} {Mathematics of computation}\ }\textbf
  {\bibinfo {volume} {51}},\ \bibinfo {pages} {699} (\bibinfo {year}
  {1988})}\BibitemShut {NoStop}%
\bibitem [{\citenamefont {Patra}\ and\ \citenamefont {Karttunen}(2006)}]{PK06}%
  \BibitemOpen
  \bibfield  {author} {\bibinfo {author} {\bibfnamefont {M.}~\bibnamefont
  {Patra}}\ and\ \bibinfo {author} {\bibfnamefont {M.}~\bibnamefont
  {Karttunen}},\ }\bibfield  {title} {\bibinfo {title} {Stencils with isotropic
  discretization error for differential operators},\ }\href@noop {} {\bibfield
  {journal} {\bibinfo  {journal} {Numerical Methods for Partial Differential
  Equations: An International Journal}\ }\textbf {\bibinfo {volume} {22}},\
  \bibinfo {pages} {936} (\bibinfo {year} {2006})}\BibitemShut {NoStop}%
\end{thebibliography}%

\end{document}